\documentclass[12pt]{article}

\usepackage[paper=a4paper,left=20mm,right=20mm,top=25mm,bottom=25mm]{geometry}
\usepackage{amsmath}
\usepackage{amsthm}
\usepackage{amssymb}

\def\ben{\begin{enumerate}} \def\een{\end{enumerate}}
\def\beq{\begin{equation}} \def\eeq{\end{equation}}
\def\beqn{\begin{equation*}} \def\eeqn{\end{equation*}}
\def\bea{\begin{eqnarray}} \def\eea{\end{eqnarray}}
\def\ba{\begin{array}} \def\ea{\end{array}}
\def\beann{\begin{eqnarray*}} \def\eeann{\end{eqnarray*}}
\def\beasn{\begin{sneqnarray}} \def\eeasn{\end{sneqnarray}}

\def\bi{\begin{itemize}} \def\ei{\end{itemize}} 
\def\bd{\begin{description}} \def\ed{\end{description}}

\def\ea{\'e}

\def\mf{\mathfrak}

\usepackage{graphicx}
\usepackage{xcolor}
\usepackage{eufrak}

\title{A contextual analysis of the early work of Andrzej Trautman and Ivor Robinson on equations of motion and gravitational radiation} 

\author
{Donald Salisbury$^{1,2}$ and Daniel Kennefick$^{3}$\\
\\
\normalsize{$^{1}$Austin College, 900 North Grand Ave, Sherman, Texas 75090, USA}\\
\normalsize{$^{2}$University of Bonn, Am Hof 1, 53225 Bonn, Germany}\\
\normalsize{$^{3}$University of Arkansas, 825 West Dickson Street, Fayetteville, AR 72701, USA}}

\begin{document}

\maketitle

\begin{abstract}
In a series of papers published in the course of his  dissertation work in the mid 1950's, Andrzej Trautman drew upon the slow motion approximation developed by his advisor Infeld,  the general covariance based strong conservation laws enunciated by Bergmann and Goldberg, the Riemann tensor attributes explored by Goldberg and related geodesic deviation exploited by Pirani, the permissible metric discontinuities identified by Lichnerowicz, O'Brien and Synge, and finally Petrov's classification of vacuum spacetimes. With several significant additions he produced a comprehensive overview of the state of research in equations of motion and gravitational waves that was presented in a widely cited series of lectures at King's College, London, in 1958. Fundamental new contributions were the
 formulation of boundary conditions representing outgoing gravitational radiation,  the deduction of its Petrov type, a covariant expression for null wave fronts, and a derivation of the correct mass loss formula due to radiation emission. Ivor Robinson had already in 1956 developed a bi-vector based  technique that had resulted in his rediscovery of exact plane gravitational wave solutions of Einstein's equations. He was the first to characterize shear-free null geodesic congruences. He and Trautman met in London in 1958, and there resulted a long-term collaboration whose initial fruits were the Robinson-Trautman metric, examples of which were exact spherical gravitational waves.

\end{abstract}

\section{Introduction}

The Polish physicist Andrzej Trautman played a key role as a transition figure from an era in the 1950's when neither the existence of gravitational waves nor the mechanism of their generation was widely recognized or understood. He made major advances in both subjects in a thesis nominally directed by Leopold Infeld, though in reality overseen by Jerzy Plebanski. His 1958 London King's College lectures, based on his Ph. D. work and the accompanying series of papers, became a sensation among the relativity congnoscenti.  It was there that he met the English mathematician Ivor Robinson who had in 1956 rediscovered the first exact solutions of Einstein's equations representing plane gravitational waves. There began a lifelong collaboration in deriving more general exact solutions that could be interpreted as representing gravitational waves. The latter part of this commentary will focus on the gravitational radiation aspect of their work in the 1950's and 1960's. They were key players in a surprisingly large international network.
\footnote{Roberto Lalli has compiled a comprehensive listing of members of the worldwide relativity network in the period around 1950 \cite{Lalli:2017aa}.} We will attempt here to isolate  some of the relevant accomplishments of this community, identify in this context the technical advances made by Trautman and Robinson, and more ambitiously try to understand how their special backgrounds, interests and talents influenced the development of gravitational radiation theory.\footnote{Daniel Kennefick has written extensively on various aspects of this history, with numerous insights that  we will have occasion to cite in this commentary \cite{Kennefick:1999aa,Kennefick:2005aa,Kennefick:2007aa,Kennefick:2017aa}. See also \cite{Blum:2018ac}.  We will be providing some additional technical detail, and will be putting some more stress on the role of Trautman, Robinson, their cited sources and their discussion partners. C. Denson Hill and Pawel Nurowski have given a good overview of some of the material presented here \cite{Hill:2017aa}. In addition Trautman \cite{Rindler:1987ab} and Robinson \cite{Penrose:1997aa} each contributed to brief overviews of each others work. Finally, Trautman himself provides some additional insights in a recent interview \cite{Trautman:2019aa}. Reminiscences of Ivor Robinson by several authors mentioned in the following can be found at the 2017 Robinson Memorial Symposium website http://artemis.austincollege.edu/acad/physics/dsalis/RobinsonMemorial/RobinsonSymposiumSchedule.html. This study was motivated in part by Salisbury's longterm relationship with Ivor and Joanna Robinson following his arrival close to the University of Texas at Dallas in 1987. Andrzej Trautman had been a frequent visitor. This review is by no means intended as an extensive historical analysis of the emergence and development of gravitational wave analysis.  }

The analysis is organized chronologically as far as possible. So as to make clear the substantial contributions of both authors  we will present sufficient technical detail to make this account accessible to non-experts in general relativity. First  we briefly review the early childhood and early education of Trautman and Robinsion. We will then give an overview of the state of affairs equation of motion and gravitational wave theory  as known by Trautman and which lead to his graduate school involvement.   As we shall see, this constitutes a not quite comprehensive prehistory of the subject. Leopold Infeld, Peter Bergmann and their collaborators are major players in this story. There follows an investigation of the links between this earlier work and Trautman's published contributions from 1956 to 1958 in the {\it Bulletin de l'Acad\'emie Polonaise des Sciences}  \cite{Trautman:1956aa,Trautman:1956ab,Trautman:1956ac,Trautman:1956ad,Trautman:1957aa,Trautman:1957ab,Trautman:1957ac,Trautman:1958aa,Trautman:1958ab,Trautman:1958ac,Trautman:1958ad}. This work was the subject of Trautman's 1958 King's College lectures.\footnote{These lectures were widely distributed as a July 1958 US Airforce publication. They were eventually published in 2002 in the Golden Oldie Series of {\it General Relativity and Gravitation} \cite{Trautman:2002aa}}   Robinson visited Trautman briefly in Warsaw in 1959 with their first resulting paper appearing in 1960 \cite{Robinson:1960aa}. They continued working together in 1961 in Syracuse with Peter Bergmann's group, eventually completing their next paper \cite{Robinson:1962aa} through correspondence while Robinson was at the University of North Carolina and Trautman at King's College. We conclude the essay with an evaluation of the gravitational radiation work performed by spouses R\'o\.za Michalska-Trautman and Joanna (Ryten) Robinson. Many of Trautman's recollections of this early work are gathered in conversations conducted in 2016 with Salisbury \cite{Trautman:2019aa}.

Andrzej Trautman was born in Warsaw on January 4, 1933. His father was a painter and worked as a high school drawing teacher. His mother Eliza, also a teacher, was of French-Polish origin, her father, Marius Andr\'e, having been a French diplomat. His formal primary school education was delayed by the outbreak of war in 1939, and then was interrupted again.\footnote{Regarding Trautman's wartime experience,  we quote from an interview conducted by Piotr Kielanowski and Agnieszka Martens \cite{Kielanowski:2007aa}. We thank Pawel Nurowski for the translation of this passage. "In the spring of 1942 my mother and I returned to Warsaw from Krzczonow.  As a result of my father's death  and of the war  we didn't have  money. My mother sold our apartment near Plac Narutowicza and we have moved to a very small apartment in Grzybowska street, formerly in the Jewish Ghetto. I started my education still at this village and then in the second half of the third grade I went to a very good private school which was at Ochota district of Warsaw conducted by  two sisters. Their names were Goldman. Then I passed to the fourth and fifth grade. When the Warsaw uprising started we were visiting
our friends in that building where our previous apartment was located.  In this neighborhood of Warsaw at that  moment there was a Nazi brigade that was led by general Kaminski, a white Russian. For the first week of the Warsaw uprising they were successively and systematically killing everybody who was thrown out from the buildings - building after building. They stopped the executions only during the eighth day of the Warsaw uprising. And it was  at the last moment, when we were standing in front of a machine gun, there  came a soldier who brought  the order to stop the executions. The Warsaw uprising began on August 1, 1944. Kaminski was a White Russian, a Nazi collaborator who was executed by the Germans at the end of August 1944 for atrocities carried out by the Russian National Liberation Army under his command.   We were then  deported to Silesia in Germany where we spent a year. My mother had to do some manual work  and I stayed at home.  That was really the worst time of my life. There was hunger. They were not feeding us properly at all and there was dirt. We were living as a group of about  30 people in one single room. Several families, you know. There were beds, one next to the other. There were not enough facilities for washing.  But it was not one of those camps - concentration camps or forced labor camps. We were deported, forcefully, during the uprising. They took us out of the house where we were living and they took us by train, not a person passenger train, a freight train. But that was a minor thing. And this ended in 1945. Since we were in Silesia in the Eastern part of Germany it was the Russians that came and we could return to Poland.  Already when we reached the town of Legnica, the Polish authorities started to operate. So we stayed with my mother for about 6 weeks there, but then we were told that we have to leave because Marshal Rokossovsky  was going to have headquarters in that town and all Poles had to go away.}

Ivor Robinson was born in Liverpool, England on October 7, 1923. His Jewish father Max Robinson, an entrepreneur, had emigrated from Russia at the age of twenty-one. Robinson received his highest degree, a B. A. in Mathematics at Cambridge University, in 1947. From 1950 to 1958 he was a Lecturer in the Department of Pure and Applied Mathematics at the University College of Wales in Aberystwyth.\footnote{Ivor Robinson passed away in 2016. A brief overview of his illustrious life and work is available here - https://news.utdallas.edu/campus-community/ivor-robinson-founding-leader-of-math-physics-depa }

\section{Research on equations of motion and gravitational radiation up through 1957 }

Trautman's research in the 1950's found inspiration and support in a rich multinational theoretical tradition. The central influence, and unifying figure, was Leopold Infeld (1898-1968). Born and educated in Poland, Infeld secured a research position with Einstein in Princeton in 1936.\footnote{See the 1941 Infeld autobiography {\it Quest} for his trials as a Jewish physicist in Poland, and the circumstances leading up to his appointment with Einstein in Princeton in 1936.}   Working with Einstein and Banesh Hoffmann (1906-1986), the three developed an iterative method, know as the EIH method, for deducing the equations of motion of structureless particles from the vacuum gravitational field equations. In 1938 Infeld joined John Lighton Synge (1987-1995) at the University of Toronto where the Dubliner Synge worked first as a lecturer from 1920 to 1925, then from 1930 to 1943 as Chair of the newly established Department of Mathematics. From 1946 to 1948 he was the head of the Mathematics Department at Carnegie Institute of Technology, after which he returned to the Dublin Institute for Advanced Studies.\footnote{See \cite{Florides:2008aa} for a discussion of Synge's impact in relativity. A strong case can be made that his group constituted the first North American relativity ``school", apart from Einstein and his co-workers at Princeton.} In 1942, upon his release from internment in Canada because he carried a German passport, Alfred Schild (1921-1977) resumed studies at the University of Toronto that had been interrupted in England. He completed his Ph. D. in 1946. Infeld directed the thesis, becoming a close friend. After a year at the Princeton Institute for Advanced Studies, he returned to Toronto for the academic year 1948-1949. In 1946 the young Briton Felix Pirani (1928-2015) began his graduate studies in relativity in Toronto.  He had received his bachelor's degree in 1948 at the University of Western Ontario, where he says ``I did well enough that I could be recommended to Infeld in Toronto". He took courses from both Infeld and Schild, receiving his Masters degree already in 1949.  Pirani went on to receive his Ph. D. under Schild at the Carnegie Institute of Technology in 1951.\footnote{See Rickles' 2011 interview with Pirani \cite{Rickles:2011aa} for more details on Pirani's relationship to Schild. Pirani's account is somewhat misleading in that he does not mention that Schild had an appointment at Carnegie in academic years 1946-48, although Schild took a leave of absence to go to Princeton with a Jewett Fellowship in 1947.  This chronology is documented in correspondence in the Alfred Schild Papers (ASP), 1915-1982, Archives of American Mathematics, Dolph Briscoe Center for American History, University of Texas at Austin.}
 Especially relevant to our story, Pirani then went to Cambridge University where he obtained a second Ph. D. under Hermann Bondi (1919-2005) in 1956. From there he went to the Institute for Advanced Study in Dublin, where  Synge had been Director since 1948. Then he returned to England in 1958 to take up a position at King's College in London where Bondi was teaching.

We will begin our overview of derivations of particle equations of motion with the 1938 EIH paper, even though, as pointed out by Havas \cite{Havas:1989aa}, there was a substantial  prehistory that EIH had largely ignored  because it relied on the material stress-energy tensor. Perhaps the most convincing was Eddington's 1923 derivation for a distribution of matter occupying a small compact spatial region \cite{Eddington:1930aa}. Eddington merely required that the finite stress-energy tensor be covariantly conserved, $T^{\mu \nu}{}_{;\nu}=0$. Then it was straightforward to show that the narrow world tube followed a geodesic.

EIH, on the other hand,  set about finding the equations of motion of interacting point singularities. The object was to show that geodesic motion obtained, and that geodesic motion was therefore a consequence of the vacuum Einstein field equations and not an independent postulate of the general theory of relativity. On its face this hope might seem absurd to a relativist - as observed by Synge in correspondence with Schild in 1948.\footnote{Schild writes on January 17, 1948, ASP, ``At the moment I am playing around with the problem of deducing the geodesic postulate for a test particle from the gravitational equations for empty space. As Einstein, Infeld, and Hoffmann have shown it should, in principle, be possible to either prove or disprove this postulate but the mathematical difficulties may be too great." Synge responds on January 20, 1948, ``I am interested to hear about the geodesic postulate, although I must say it is difficult to understand. How can a line of singularities be a geodesic? This point has puzzled me for a long time, and has checked me from accepting the basic ideas of Einstein, Infeld and Hoffmann. Perhaps you have a clearer way of viewing the situation."} It is questionable whether  such motion could be meaningfully assigned even to a lone Schwarzschild singularity.\footnote{It is interesting - and relevant to our later story - that Ivor Robinson \cite{Robinson:1985ab} was led to his rediscovery of the plane wave solution of Einstein's equations through his attempt to boost a point mass in special relativity to the speed of light.} A related objection has to do with absence of any preferred coordinate system in general relativity. This was stated succinctly by Bergmann in 1950 in a review he wrote of a paper submission by Infeld and Scheidegger \cite{Infeld:1951aa}, to which we shall return. According to Bergmann, it would be sensible to introduce approximately flat Minkowski coordinates only near spatial infinity.\footnote{Syracuse University Bergmann Archive (SUBA)``I agree that any kind of motion in terms of the coordinates can be produced by an appropriate choice of coordinates, and that, thus, relativistic effects can be thrown from the equations (or rather expressions ) for the world lines of the particles into the form of the metric. But this statement holds also for the non-relativistic motion. We could require that the two particles of a two-body system remain permanently at rest.  Naturally, the required coordinate transformation is rather brutal. I would say that the reasonable choice of a coordinate system should exhibit the so- called relativistic effects uniquely in the form of the world lines traversed by the particles. And such a reasonable choice includes flatness of the space at (spatial) infinity, and the absence of secular terms in the metric in the case of the two-body problem, except possibly at such high approximations that radiation damping would become observable. `Reasonable' restrictions do not include coordinate condition (in the customary sense). "Letter from Bergmann to Coxeter, June 20, 1950. }  The freedom of choice in coordinates played an essential role in EIH's original paper, and in subsequent collaborations of Einstein and Infeld \cite{Einstein:1940aa, Einstein:1949aa}. In 1938 they were able to ignore the spatial region near the particle singularity, presumably far enough away such that the metric differed only slightly from the Minkowski metric, i.e., $g_{\mu \nu} = \eta_{\mu \nu} + h_{\mu \nu}$ where $\left|h_{\mu \nu}\right| << 1$. This was a consequence of their discovery that with appropriate coordinate conditions  four of the vacuum Einstein's field equations took the form, where $\gamma_{\mu \nu} := h_{\mu \nu} - \frac{1}{2} \eta_{\mu \nu} \eta^{\sigma \rho}h_{\sigma \rho}$,
\beq
\left(\gamma_{0 a,b} - \gamma_{0 b,a}\right)_{,b} = \gamma_{00,0a} + 2\Lambda_{0 a}, \label{ein1}
\eeq
and
\beq
\left(\gamma_{c a,b} - \gamma_{c b,a}\right)_{,b} = 2\Lambda_{c a}, \label{ein2}
\eeq
where $\Lambda_{\mu a}$ is non-linear in $h_{\mu \nu}$. The anti-symmetry of the  left  hand sides under the interchange $a \leftrightarrow b$ implies that these terms may be written as the $a$'th component of a curl, namely,
\beq
\left(\gamma_{\mu a,b} - \gamma_{\mu b,a}\right)_{,b} = \epsilon^{a b c} \partial_b \epsilon^{c d e} \gamma_{\mu d, e}. \label{anti}
\eeq
Thus it follows by Stokes theorem that if this curl is integrated over any closed surface where the integrands are well-defined, then since the boundary of the two-dimensional surface vanishes,  the surface integrals themselves must vanish. This is the key to the entire enterprise since we deduce that the following integrals must vanish,
\beq
\oint \left( \gamma_{00,0a} + 2\Lambda_{0 a} \right) n^a dS = 0, \label{ein1int}
\eeq
and
\beq
\oint \Lambda_{c a} n^a dS = 0, \label{ein2int}
\eeq
where $n^a dS$ is the spatial surface area element.

Equations (\ref{ein1int}) and (\ref{ein2int}), combined with the third vacuum field equation
\beq
R_{00} = -\frac{1}{2} \gamma_{00, aa} + \Lambda_{00} = 0, \label{ein3}
\eeq
where $\Lambda_{00}$ is also non-linear in $h_{\mu \nu}$ form the basis of the EIH approximation procedure. They proposed an iterative solution in powers of a small parameter $\lambda$. This parameter is conceived as a measure of the assumed relatively slow rate of change of the metric with respect to $x^0$ as opposed to $x^a$, i.e., it is assumed that $\frac{\partial }{\partial x^0} << \frac{\partial }{\partial x^a}$. On the other hand the rate of change relative to a new coordinate $\tau := \lambda x^0$ (also with dimension of length) is assumed to be comparable to the rate of change with respect to $x^a$, implying that $\lambda << 1$. It follows from this assumption that if one were to take into account the stress-energy  tensor $T_{\mu \nu}$ of matter as the source of curvature, then the components would satisfy the inequalities $\left|T_{ab}\right| \approx \lambda \left|T_{0a}\right| \approx \lambda^2 \left|T_{00}\right|$.  In addition,  as noted in \cite{Einstein:1949aa}, while $T_{00}$ is approximatelly the particle source energy density, one can ``by pure convention"  take the leading contribution to $\gamma_{00}$ to be of order $\lambda^2$.\footnote{In \cite{Einstein:1938aa}, p. 73, the observation is made that this choice follows from the  argument that due to energy conservation the potential energy per unit mass "is of the same order as the square of the velocity". } Now since the linearized Einstein equations take the form
\beq
\gamma_{\mu \nu, \rho \sigma} \eta^{\rho \sigma} = -\frac{16 G}{c^4}T_{\mu \nu}, \label{linein}
\eeq
 we deduce  the additional leading behaviors $\gamma_{0a} \approx \lambda^3$ and $\gamma_{ab} \approx \lambda^4$. Then
 recognizing that only  second time derivatives appear in (\ref{linein}) we are led to the following expansions to be substituted into the field equations (\ref{ein1int}), (\ref{ein2int}), and (\ref{ein3}),
$$
\gamma_{00} = \lambda^2 {}_{2}\!\gamma_{00} + \lambda^4 {}_{4}\!\gamma_{00} + \dots,
$$
$$
\gamma_{0a} = \lambda^3 {}_{3}\!\gamma_{0a} + \lambda^5 {}_{5}\!\gamma_{0a} + \dots,
$$
and
$$
\gamma_{ab} = \lambda^4 {}_{4}\!\gamma_{ab} + \lambda^6 {}_{6}\!\gamma_{ab} + \dots
$$

Finally we are prepared to appreciate the value of the surface integrals in (\ref{ein1int}) and (\ref{ein2int}). We learn from (\ref{ein3}) that the lowest order equation for $\gamma_{00}$ is the Laplace equation
\beq
{}_{2}\!\gamma_{00, aa} = 0, \label{laplace}
\eeq
recalling that this is required to hold in vacuum.
Assuming that we have two interacting particles, this provides the occasion to introduce singular sources at $\xi_1^a(x^0)$ and $\xi_2^a(x^0)$ at time $x^0$ resulting in  the following solutions of (\ref{laplace}):
\beq
{}_{2}\!\gamma_{00} = -4 m_1(x^0) \left[ \left(x^a-\xi_1^a(x^0)  \right) \left(x^a-\xi_1^a(x^0)  \right)\right]^{-1/2} -4 m_2(x^0) \left[ \left(x^a-\xi_2^a(x^0)  \right) \left(x^a-\xi_2^a(x^0)  \right)\right]^{-1/2}.
\eeq
It is apparent that when one now performs surface integrals with only particle number $n$ in the interior, the only terms that will survive are those that vary as $\left[ \left(x^a-\xi^a_n(x^0)  \right) \left(x^a-\xi^a_n(x^0)  \right)\right]^{-1/2}$, and they will depend on the $\xi^a_n(x^0)$, as well as on time derivatives of these particle positions as is apparent with the time derivative of $\gamma_{00}$ that appears in (\ref{ein1int}), i.e., the time derivatives arise due to the time derivatives that appear in the field equations.

With this presumed Newtonian form one begins the iterative expansions with the remaining field equations (\ref{ein1}) and (\ref{ein2}). It turns out that these become, at each order of $\lambda$, Poisson equations. But there remains crucial consistency conditions that must be fulfilled. The antisymmetry of the left hand sides of (\ref{ein1}) and (\ref{ein2}) imply that the right hand sides must obey the divergence conditions
\beq
\gamma_{00,0aa} + 2\Lambda_{0 a,a} = 0,
\eeq
and
\beq
2\Lambda_{c a, a} = 0.
\eeq
Not incidentally, these conditions must also be satisfied to insure that the surface integrals (\ref{ein1int}) and (\ref{ein2int}) are independent of the particular surfaces chosen to contain the particle sources.
Einstein and Infeld developed an increasingly elegant means for satisfying these conditions in 1940 \cite{Einstein:1940aa} and 1949 \cite{Einstein:1949aa}.\footnote{Perhaps not surprisingly, Bergmann included a discussion of the earlier formalism in his 1942 textbook \cite{Bergmann:1942aa}. The forward to the book was contributed by Einstein, with whom he had worked at Princeton from 1936 to 1941. } In the latter paper they described how one could satisfy these conditions at each iterative step by adding fictitious monopole and dipole terms to the solutions of the various Poisson equations. Having decided to conclude the iteration at some power $n$ of $\lambda$, one then was to require that the three integrals  (\ref{ein2int}) were to vanish,
\beq
\oint \sum_{k=2}^n \lambda^k {}_k\!{\Lambda_{c a}} n^a dS = 0. \label{eqmot}
\eeq
As we shall see, these are the particle equations of motion!  ( $\lambda$ is eliminated by absorbing it into the definitions of the masses and the time $\tau$.)

Let us now return to the key discovery that made possible the deduction of the equations of motion, the isolation of the anti-symmetric terms (\ref{anti}) in the field equations. As EIH observed in 1938, it is possible to obtain these terms by imposing coordinate conditions. That this was possible was a consequence of the fact that four of the Einstein field equations do not contain second time derivatives of the metric, and therefore one is dealing with ``an over determined system of equations". They observed that ``the overdetermination is responsible for the existence of equations of motion".\footnote{ \cite{Einstein:1938aa}, p. 65} The overdetermination is of course a consequence of the general covariance of Einstein's field equations. Two years later Einstein and Infeld showed \cite{Einstein:1940aa} that the antisymmetric contribution could be isolated without imposing coordinate conditions, but neither there or in their final collaboration in 1949 \cite{Einstein:1949aa} did they pursue a direct link between the appearance of the antisymmetric contribution and general covariance. A first step in this direction was made by Peter Bergmann in 1949 \cite{Bergmann:1949aa}. Bergmann had actually joined Einstein in Princeton at the same time as Infeld, and he was intimately familiar with the EIH procedure. It could very well have been his principle intention in this first of his series of papers on general covariance to elucidate the apparent connection to EIH. Bergmann showed how a generalization of the contracted Bianchi identities arose in any generally covariant field theory - and consequently non-linear - field theory.\footnote{That there existed a link between the general covariance of Einstein's field equations and the contracted Bianchi identities was of course well known, to Einstein in particular \cite{Einstein:1918ae}. The analysis can be traced to Felix Klein \cite{Klein:1918aa}, and was summarized by Pauli \cite{Pauli:1921aa}\cite{Pauli:2000aa}} 

Bergmann lets $y_A$ represent a generic field and he assumes that under a general infinitesimal coordinate transformation $x'^\mu = x^\mu + \xi^\mu(x)$ its variation is $\bar \delta y_A = F_{A \mu}{}^{B \nu} y_B \xi^\mu_{,\nu}  - y_{A,\nu} \xi^\nu$ where the $F_{A \mu}{}^{B \nu} $ are constants.\footnote{ The symbol $ \bar \delta $ represents the Lie derivative, in this case relative to the infinitesimal displacement $-\xi^\mu(x)$.} Then assuming that the action is invariant under these transformations one obtains the identities
\beq
\left( F_{A \mu}{}^{B \nu} y_B L^A \right)_{, \nu} + y_{A,\mu} L^A \equiv 0, \label{Bianchi}
\eeq
where $L^A := \frac{\partial L}{\partial y_A} - \left(\frac{\partial L}{\partial y_{A, \mu}}  \right)_{, \mu} = 0$ are the field equations. As noted by Bergmann, in the case of general relativity these are the contracted Bianchi identities.

 Now representing as above the contributions to various orders of $\lambda$ with a left suffix, Bergmann isolated the first nontrivial linear term on the left hand side of the field equations, in the second order, as
$$
 L^{AB ab} {}_2 y_{B, ab} = - {}_2 L^A,
$$
where the $ L^{AB ab} := - \frac{\partial^2 L}{\partial y_{A,a} \partial y_{B,b}} $ are constants. But the appropriate linear combination of the left hand side satisfies the identity that follows from (\ref{Bianchi}), namely
$$
F_{A \mu}{}^{C c} {}_0y_C \left(L^{AB ab} {}_2 y_{B, ab} \right)_{,c} \equiv 0,
$$
where the ${}_0 y_C$ are constants.  (For the case of general relativity they would be the Minkowski metric.) This means that even though we do introduce a singular field,  the ${}_2 y_B$ can be chosen arbitrarily near the region where the singularity occurs. As a consequence this expression will be well-defined everywhere and it has vanishing divergence. Thus one can perform a surface integral of this linear combination of the field equations. Bergmann  therefore made the direct link between general covariance and the ability to express sums of the terms of the field equations as a surface integral.

This work was part of the inspiration for Joshua Goldberg's Ph. D. thesis, written under Bergmann's direction. He published his initial results in 1953 \cite{Goldberg:1953aa}. Goldberg formally established the desired link between the antisymmetric contributions to the field equations and the particle equations of motion. He showed that from the identities (\ref{Bianchi}) there follows the `strong' (i. e., identically satisfied) conservation laws\footnote{We employ the subscript ``s"  for `strong'}, thus distinguishing this object from the material stress-energy.
\beq
T_{s \,\mu}{}^\nu{}_{,\nu} \equiv 0, \label{strong}
\eeq
 where
\beq
T_{s \,\mu}{}^\nu := - F_{A \mu}{}^{B \nu} y_B L^A + t_\mu{}^\nu,  \label{Ts}
\eeq
and
\beq
t_\mu{}^\nu :=- \delta_\mu{}^\nu L+ \frac{\partial L}{\partial y_{A,\nu}}y_{A,\mu}   \label{pseudo}
\eeq
 is the `pseudo stress energy tensor, `pseudo' because it does not transform under arbitrary general coordinate transformations as a tensor.
The identical vanishing of the divergence implies that $T_{s \,\mu}{}^\nu $ may be written in terms of an anti-symmetric `superpotential' $U_\mu{}^{[\nu \sigma]}{}_{, \sigma}$, i.e.
\beq
T_{s \,\mu}{}^\nu = U_\mu{}^{[\nu \sigma]}{}_{, \sigma}. \label{U}
\eeq
Then according to (\ref{Ts}) and (\ref{U}) we can write certain linear combinations of the field equations in terms of the superpotentials,
\beq
F_{A \mu}{}^{B \nu} y_B L^A \equiv U_\mu{}^{[\sigma \nu] {}_{, \sigma}}+t_\mu{}^\nu = 0. \label{superident}
\eeq
The first term, anti-symmetric in the  upper spatial indices, is the term Goldberg had been seeking. Curiously, although he did cite the use of this tensor by \cite{Freud:1939aa} he did not delve more deeply into its history. With its aid he could then form surface integrals over  regions enclosing singularities, and carry out the EIH slow motion iterative derivation of the particle equations of motion.\footnote{Goldberg informed Salisbury that he did get to describe this work to Einstein in Princeton in 1954. He reported that Einstein showed little interest in the strong conservation laws - most certainly because he was already aware of them. And he seemed even less interested in equations of motion. He was more interested in seeing exact gravitational wave solutions. Robinson's rediscovery  of them would only come two years later. See \cite{Salisbury:2023ab} for more details of the Goldberg-Einstein conversation.}

We move on now to procedures for deducing equations of motion which are focused on the stress-energy tensor rather than the contracted Bianchi identities. Infeld first turned his attention to this problem in 1954 \cite{Infeld:1954aa}.\footnote{Trautman wrote in 1978 a brief overview of Infeld's work on the equations of motion \cite{Trautman:1978aa} in a collection and commemoration  edited by Infeld's son Eryk Infeld \cite{Infeld:1978aa}. The volume also contains the same brief biography and the complete annotated Infeld bibliography that first appeared in 1970 \cite{Infeld:1970aa}.  } Infeld began this paper with the observation that Fock had almost simultaneously and independently of EIH developed an approximation technique based on the vanishing covariant divergence of the stress-energy tensor of a material source. He wrote that this difference was unimportant ideologically since the singularities in the EIH method were just useful mathematical abstractions\footnote{It is perhaps no coincidence that likely one of the first topological objections to the use of the singularity assumption, by Dirac \cite{Dirac:1964ab}, was published after the discovery of quasars - although his talk was given in Warsaw in 1962. }, but he did take issue with Fock's insistence that there existed preferred coordinate conditions. Also, Fock dealt with continuous distributions of matter - without singularities. Infeld proposed to take the stress-energy tensor of matter into account, but to model the stress-energy with singularities with Dirac delta functions.\footnote{The Dirac delta functions had in fact been used previously by the Polish physicist Myron Mathisson \cite{Mathisson:1931ab}. Indeed, in 1937 Mathisson deduced equations of motion of spinning particles in a gravitational field using a precurser of genuine distributions \cite{Mathisson:1937aa}. The resulting equations are now referred to as  the Mathisson-Papapetrou equations. An English translation of this article has appeared as a Golden Oldie \cite{Mathisson:2010aa}, with an editorial note by Trautman \cite{Trautman:2010ac}.  See the article by Sauer and Trautman for an account of Mathisson's life and work - and in particular his interaction with Einstein \cite{Sauer:2008aa}. Mathisson also features in the epilogue of Havas' early history of the problem of motion in general relativity , lamenting that ``Infeld totally ignored Mathisson's work", and opining that ``Infeld made an enormous contribution by establishing his own school of relativists in Poland. On the other hand, his dominion of it ... had a retarding effect on the development of approaches other than his own, and there was little awareness in Poland of work other than that of Einstein and his collaborators and of Fock, on the problem of motion."\cite{Havas:1989aa}, p. 267. The present overview of international influences on Trautman does tend to belie this notion. It is interesting, though, that Infeld might have felt some animosity toward Mathisson, evidenced for example  in a letter from Opechowski to Schild, ASP, dated October 8, 1949, letting Schild know that he may keep the Lubanski article he had lent him, and then inquiring ``what the possible objections are against Mathisson's method ... You for instance did not even know that this work had been done, and Infeld did never take the trouble of glancing through those papers. I understand that Infeld and Mathisson did not like each other." } Two years later, in collaboration with Plebanski \cite{Infeld:1956ab}, he showed how calculations could be simplified by introducing 'good' delta functions, $\int d^3x \hat \delta^3 x (x)/r^p = 0$, with $p$ a positive interger.\footnote{See also \cite{Infeld:1957aa} for more details.} Since integrations over singular regions were now permitted, it was no longer necessary, at least from the point of view of particle equations of motion, to work with surface integrals. Nevertheless, the Infeld-Plebanski procedure continued to be based on the underlying general covariance of Einstein's field equations, now with explicit stress-energy sources,
$$
G_{\mu \nu} = -  \frac{8 \pi  G}{c^2}T_{\mu \nu},
$$
with stress-energy of pole particles given by
$$
T^{\mu \nu} = \sum_i m_i \dot \xi^\mu_i  \dot \xi^\nu_i \hat \delta^3 \left(x^a - \xi^a(x^0)  \right),
$$
and where $\dot \xi^\mu:= \frac{d\xi^\mu}{dx^0}$ and $ \xi^0_i = x^0$.
The divergence of the left hand side vanishes identically (the contracted Bianchi identity following from the general covariance), and consequently to be consistent the covariant divergence of the stress-energy must vanish,
\beq
T^{\mu \nu}{}_{; \nu} = T^{\mu \nu}{}_{, \nu} + \Gamma^\mu_{\sigma \nu} T^{\sigma \nu} + \Gamma^\nu_{\sigma \nu} T^{\mu \sigma} = 0. \label{div}
\eeq
Using the same slow variation approximation procedure as above, Infeld and Plebanski imposed these conditions iteratively  in powers of $\lambda$. Integrations over regions enclosing individual poles yield the individual particle equations of motion. The computational task is in fact somewhat simplified by the smaller number of contributing components of $h_{\mu \nu}$ that appear in (\ref{div}) to each order.

It is perhaps not surprising that the shift in emphasis from the gravitational field to the particles ushered in an attempt to exclude the fields entirely from the formalism.  This approach had been pioneered by Lorentz and Droste \cite{Lorentz:1917aa}, with later contributions by Fichtenholz \cite{Fichtenholz:1950aa}. Infeld addressed this problem in 1957, displaying the two-particle Langrangian that described the first post-Newtonian approximation, but not mentioning the long history. The resulting equations of motion for the two-particle system agreed with the sixth order results of the original EIH calculation. These equations were actually solved by Robertson in 1938 \cite{Robertson:1938aa}.

This brings us finally to the question whether either the EIH or the Infeld-Plebanski procedure can take the production or absorption of gravitational waves into account, with a corresponding back reaction. Indeed, whether gravitational waves exist or not was vigorously debated during this period prior to 1958. It was clear already in 1938 that since the EIH procedure was time symmetric, as originally formulated, propagating waves could not arise. The gravitational fields were equal admixtures of advanced and retarded waves. This was well understood, and illustrated by Infeld in 1938 \cite{Infeld:1938aa} with the analogous $1/c$ expansion applied to electrodynamics. He showed that when odd powers of $\lambda$ were retained in $\gamma_{00}$ and $\gamma_{ab}$, and even powers in $\gamma_{0a}$, then it was possible to get either advanced or retarded wave solutions. It turns out that through the sixth order in $\lambda$ there is no alteration in the equations of motion. Infeld showed that radiation effects could only appear at the ninth order - and then according to Robertson's calculation any average secular change in particle orbits would appear only at the eleventh order. He concluded that ``the result shows the astonishingly small role played by the gravitational radiation in the motion of double stars."\footnote{\cite{Infeld:1938aa}, p. 841} Infeld continued to address the question of gravitational wave production and back reaction with his Toronto students. He and Wallace achieved a promising first step in 1940 \cite{Infeld:1940aa} when they showed that Maxwell's equations in flat space could be put into  the form of  vanishing spatial divergences - outside the location of a singular charged point source. The condition that the surface integral enclosing the source was required to vanish yielded in a much more straightforward manner than Dirac's procedure \cite{Dirac:1938aa} -  the electromagnetic back reaction on an accelerating electric charge.

We have already referenced the 1949 paper by Infeld and Schild in which they described a limiting procedure for obtaining the geodesic motion of a test particle by considering the limit of a mass $m$-dependent spacetime. As motivation for this work they cited a fact that was drawn to their attention by Wheeler, that the EIH procedure is well suited to the case of slowly varying fields, but ``is inadequate to our present problem: the motion of a small mass in an arbitrarily strong field".\footnote{\cite{Infeld:1949aa}, p. 409} Although they did not address the problem of back reaction in this paper, the subject did emerge in subsequent correspondence with Wheeler. Replying to a letter from Schild, Wheeler writes ``About the problem of gravitational radiative reaction, I believe the question is a very interesting matter of principle which deserves a careful analysis before one attempts by purely mathematical means to arrive at the result. It seems to me in fact, doubtful that one can in principle express the gravitational radiative reaction on a particle in terms only of the motion of that particle and the properties of the metric - background metric - along the path of that particle. I have been thinking about this problem from time to time and if I have any further comments, I will write to you and Infeld."\footnote{ASP, letter dated May 20, 1949} Schild responds ``Infeld and I agree with your remarks. We decided some time ago that an extension of the geodesic equation to include gravitation radiation reaction does not have too much meaning. It seems that, for the case of particles of non-vanishing mass, the essential problem is not the geodesic equation as such (or any similar equation), but rather the appropriate definition of a background field. We have done some rather sketchy and incomplete work on the problem, which was interrupted when Infeld left for Europe in April and I moved to Pittsburgh. We believe that it has no meaning to talk rigorously about the geodesic motion of a particle of finite mass - but that one can talk about geodesic motion in connection with an approximation procedure. In the case of the Einstein-Infeld-Hoffmann method it seems possible to define in an invariant manner a `background field' for which the motion is along a geodesic - but only up tho the 12th approximation - in the 14th approx. the method breaks down for apparently good physical reasons (at this stage, the background field enters in a non-linear manner). We would be interested to hear your comments on these remarks or on the problem of gravitational radiation."\footnote{ASP, letter dated June 20, 1949}

In 1951 Infeld with his student Scheidegger  \cite{Infeld:1951aa} investigated whether gravitational wave back reaction was a coordinate-dependent phenomena - concluding that indeed it was. It should be stressed that what they referred to as radiative were any effects that resulted from the odd or even powers of $\lambda$ that had been omitted in the original EIH program. Scheidegger had already published the earlier claim from his thesis that Hu's \cite{Hu:1947aa}\footnote{Ning Hu earned a doctoral degree at the California Institute of Technology in 1943, arriving in 1940 from Beijing. From 1943 to 1945 he worked with Wolfgang Pauli at the Princeton Institute for Advanced Studies before going to the Dublin Institute for Advanced Studies.} surprising deduction that the energy of a binary system actually increased over time was a coordinate effect, concluding that ``Two moving bodies, such as double stars, are in a secularly stable movement, they do not radiate gravitational waves and there exists no slowing down of the motion by radiation damping."\footnote{\cite{Scheidegger:1951aa}, p. 223} It is noteworthy that Scheidegger did not reject the notion that non-linear gravitational waves could exist. ``If it is possible to solve the gravitational field equations by an entirely different method than the quasi-static approach of Einstein, Infeld, and Hoffmann, then it still might be that there exists a non-linear phenomenon which one could call a {\it wave}."\footnote{\cite{Scheidegger:1951aa}, p. 223}. Infeld and Scheidegger cited as supporting evidence of their claim the 1949 work by Bergmann and Bruning \cite{Bergmann:1949ab}, the continuing development of a generally covariant Hamiltonian formulation of general relativity. Bergmann refereed their paper, and his comments are illuminating. He did approve the paper for publication, but he noted some disagreements, and wrote that `` Unless, as a result of my comments, Professor Infeld or Dr. Scheigegger should decide to modify the paper, I should be perfectly happy to have it appear in its present form. My comments refer to a general point of view and also to a particular result claimed by the authors." Regarding gravitational radiation he wrote "The assertion that the gravitational radiation, what there may be introduced of it, can be transformed away. Now I would like to say that the radiation indicated by eqs. (3.1) of the present paper, quoted from ref. (5)\footnote{The reference is to \cite{Infeld:1938aa}}, can in fact be transformed away and represents nothing but a vibrating coordinate system. However, it is well known from the so-called linear approximation that only the transverse-transverse waves are real, as indicated, for instance, on p. 188, eqs. (12.38) and (12.39) of my book (Introduction to the Theory of Relativity \footnote{\cite{Bergmann:1942aa}}). Therefore, we should look for physically significant waves only in an approximation that is consistent with the transverse-transverse character of the only real gravitational waves. Let us not forget that the linear approximation is undoubtedly valid at spatial infinity. Now, returning to reference (5), we note that such a transverse-transverse wave will be induced only in an approximation higher by $c^{-2}$ than the one in which the transverse-longitudinal wave and the purely longitudinal waves first appear. This approximation has not been studied in (5). Whether or not a real gravitational wave is produced by a double-star system can, in my opinion be decided only if we go to that approximation in which $\Lambda_{mn}$ no longer vanishes. Should it turn out that such a wave is produced, it will make a real difference whether we adopt stationary solutions (even orders of approximation only) or whether we accept retarded solutions corresponding to mixed orders. That real difference will, however, appear only in an exceedingly high approximation, that one in which the squared amplitudes of the transverse-transverse waves first appear. Scheidegger's results, on the other hand, appear open to the objection that in his approximation no real waves whatsoever appear. Naturally, a flat fictitious wave is equivalent to a fictitious progressive wave. Both can be got rid of."\footnote{SUBA, Letter from Bergmann to Coxeter, dated June 20, 1950} \footnote{There is an interesting addendum to this review. Infeld had made the decision to return to Poland without informing Scheidegger, and he wrote to Bergmann on August 29, 1950, SUBA, seeking advice on what to do with the paper, writing ``I had a talk with Dean Beatty; he does not know himself what the score is. Dr. Infeld just sent him a wire from England that he won't be back." "Under these circumstances I thought it best to suggest to Dr. Coxeter to print the paper on grav. rad. and motion as it stands, for I cannot change it very well without the consent of the co-author, and Dr. Infeld might never be heard of again from behind the iron curtain. Who knows? - I shall write a little sequel on transverse-transverse waves myself at a later date. I hope this meets with your approval, too. Bergmann replied on October 6, 1950 ``"Thanks for the news about Infeld. Coxeter wrote me a very nice letter, which says that Infeld also wishes to go through with the publication as the paper stands. I think that your idea is very sound, to leave any further work to a future paper." } Scheidegger did indeed follow up in 1951 on Bergmann's observation with a paper on gravitational transverse-transverse waves \cite{Scheidegger:1951ab}. He claimed to show that in the context of the EIH approach the transverse-transverse waves could indeed be eliminated through appropriate coordinate transformations! Goldberg also followed up on Bergmann's criticism, publishing his groundbreaking paper on gravitational radiation in 1955 \cite{Goldberg:1955aa}.\footnote{The paper was first submitted in 1953 while Goldberg was employed at the Armour Research Laboratory. (He was able to devote roughly half his time there to his own research according to a private communication.)} Goldberg showed that it was indeed possible through a coordinate transformation to eliminate the radiation terms ${}_i\!\gamma_{00}$ for odd $i$ and ${}_j\!\gamma_{0a}$ for even $j$. Furthermore it was always possible through a coordinate transformation to remove the trace $\delta^{ab}{}_k\!\gamma_{ab}$ for odd $k$. Goldberg's crucial observation was that the true indicator of coordinate-independent gravitational waves was the presence of non-vanishing Riemann curvature. Scheidegger's error was that he had not realized that the removal of the trace-free transverse-transverse  field at a given order introduced in higher order the gauge function that needed to be taken into account in computing this curvature.

Goldberg also followed Infeld's lead \cite{Infeld:1938aa} in performing the slow motion expansion in every order so as to include gravitational radiation. He argued, however, that the slow motion EIH method could not be employed as additional knowledge was required about the motion of singularities. Pertinent for our later story is nevertheless his detailed construction, appearing in Appendices, of the superpotential $U_\mu^{[\nu \sigma]}$ and pseudotensor $t_\mu{}^\nu$. Goldberg deduced from the resulting equation (\ref{superident}), applying the EIH procedure, that if the radiation matching terms appeared in the fifth order they  contributed to the equations of motion in the ninth order, and to radiation itself in the tenth order.

The relation of the Riemann tensor to the deviation of geodesics was first deduced in 1926 by Levi-Civita \cite{Levi-Civita:1926aa} and independently the same year by Synge \cite{Synge:1926ab}. Pirani learned of this physical significance in the course on general relativity he took with Schild in Toronto in 1948-49.\footnote{Schild referred to this course in a letter of recommendation, ASP, he wrote for Pirani on February 28, 1955. } Indeed, Schild and Synge were at that time preparing their standard text on tensor calculus in which geodesic deviation played a featured role \cite{Schild:1949ab}. Thus Pirani was well prepared to recognize the significance of the Riemann tensor in relation to the detection of gravitational waves. But first in 1956, while working as a post-doctoral researcher under Synge in Dublin, he published a paper on the physical significance of the Riemann tensor. In this paper he was one of the first to use Fermi-Walker \cite{Walker:1933aa} \cite{Fermi:1922ab,Fermi:1922aa} transported orthonormal tetrads. Already, in a paper presented at the meeting in Bern in 1955 \cite{Pirani:1956aa}, Pirani had made the case that observable physical effects could be meaningfully defined relative to local inertial frames of reference, in particular with respect to non-rotating Fermi transported coordinates. The emphasis in this article was on geodesic precession - related to the motion of particles with spin. He continued this line of thought in 1956 with a paper on the physical significance of the Riemann tensor \cite{Pirani:1956aa}.\footnote{This paper was reprinted as a Golden Oldie in 2009 \cite{Pirani:2009aa}. Trautman added editorial comments, a brief Pirani autobiography and his own observations on Pirani's career \cite{Trautman:2009aa}. He makes a reference to a remark in the preface of Synge's relativity text, which merits repeating here, ``Dr. Pirani introduced me to the transport law of Fermi which plays an important part in the book, and my attempt to turn Riemannian geometry into observational physics (measure the Riemann tensor!) originated largely in discussions with him ...\cite{Synge:1960ad}, pp. X-XI} We will employ Pirani's notation, where in both Greek and Latin suffix alphabets, only letters in the second half are used for tensor and connection indices. Letters from the first half are used to label tetrad quantities. A tetrad with label $\alpha$ is represented by $\lambda_\alpha^\mu$. Unprimed numbers refer to tetrad labels and primed numbers to tensor indices. Then the Fermi-Walker coordinates $X^a$ satisfy the geodesic deviation relation
$$
\frac{d^2 X^a}{d \tau^2} + R^a{}_{0 b 0} X^b = 0,
$$
where $\tau$ is the proper time along the observer world line. Pirani also briefly addressed in this paper what discontinuities could arise in the metric when employing Fermi-Walker coordinates. Both these observations played key roles in the subsequent paper on an invariant formulation of gravitational radiation theory, published in 1957 \cite{Pirani:1957aa}. This work had already been submitted prior to the Chapel Hill Conference on the Role of Gravitation in Physics that took place in January, 1957. Pirani presented the results of the 1957 paper there \cite{Pirani:2011aa}.\footnote{The Pirani contribution did not appear among the collection published in the Reviews of Modern Physics in July, 1957. It was contained in a report edited by C\ea cile DeWitt for the conference funding agency, the Wright Air Development Center, and is available as a government document. Dean Rickles and DeWitt have recently published this report with some small corrections and annotations. One wonders what to make of a remark Pirani made in a note to Schild, ASP, on November 30, 1957, in which he states ``after reading July Rev. Mod. Phys. I wonder whether I really do GR or something else." Pirani also informed Schild in this note that he would be visiting Infeld in Warsaw in December, 1957, ``staying 2-3 weeks to talk to people and see what work they are doing there."} It was in the discussion period following Pirani's talk that Bondi asked ``Can one construct in this way an absorber for gravitational energy by inserting a
$\frac{d \eta^a}{d \tau}$ term, to learn what part of the Riemann tensor would be the energy producing one, because it is that part that we want to isolate to study gravitational waves." Pirani responded ``I have not put in an absorption term, but I have put in a `spring'. You can invent a system with such a term quite easily."\footnote{\cite{Pirani:2011aa}, p. 142} (The separation coordinate $\eta^a$ is represented above by $X^a$.)

We get some insight into Pirani's developing ideas and challenges in early 1956 in correspondence with Schild. Pirani writes ``Bondi and I have been working on gravitational waves for some months without making an awful lot of progress... The underlying physical idea about vielbeins is one which really comes from Synge. At any point the timelike vector of a vierbein is to be identified as the 4-velocity of an observer using the 3 spacelike vectors as a system of local Cartesian coordinate axes. Then out of any tensor one can construct scalars which are the physical components of the tensor as measured by that observer using those axes... In effect, deviations from flatness will appear not in terms of the Christoffel symbols but in terms of the vierbein derivatives (Ricci rotation coefficients)... The problem of gravitational waves seems to me to be divided into two parts: in the first place there is the difficulty that the pseudotensor $t^\mu_\nu$ which comes into
$\frac{\partial}{\partial x^\nu}\left( T^\mu_\nu + t^\mu_\nu \right) = 0$
is a pseudo-tensor and not a tensor. This makes it impossible to attach any reasonable physical significance to calculations of the gravitational field energy, etc, although lots of people do this in the weak field approximation (eg. Joshua Goldberg in Phys. Rev. last October  and Landau and Lifshitz  Theory of Fields)... Because the covariant conservation law is a \underline{tensor} divergence $T^{\mu \nu}{}_{;\nu} = 0$ rather than a \underline{vector} divergence law $P^\mu_{;\mu} = 0$ one cannot construct invariant integral conservation laws for matter as one could if one had a vector which could be converted into a surface integral by using the Gauss-Green formula ... This is one basic difficulty. Another one is just the actual \underline{definition} of gravitational waves or radiation. I have made some progress with this problem of definition, but the conservation law problem is holding everything up..."\footnote{ASP, Undated letter from from Pirani to Schild, but written prior to May 24, 1956, which is the date of Schild's response in which he writes ``so sorry to have been so long in answering your letter about gravitational waves." } The progress that Pirani reports with the definition of gravitational waves has to do with the observation that ``the gravitational field is characterized by the full Riemannian tensor in empty space", and this does follow naturally from Goldberg's result which does begin to take the non-linearity of the field equations into account through the EIH iterative approach. However, it had not yet been possible to establish a criterion that would be valid in an exact sense. Pirani writes that ``one can get an idea toward the definition of grav. waves by considering the possible canonical forms of the Riemann tensor. The idea is to define a sort of principal directions for the Riemann tensor analogous to the Ricci principal directions. I won't go into details but the essential result is that one can define such directions and under certain circumstances some of them can become null directions. It turns out that in this case the canonical form contains extra terms; these extra terms come in just the components that Synge-O'Brien or Lichnerowicz continuity conditions allow discontinuities in, across a null 3-surface. This is pretty condensed, but what I want to do is to associate null directions of the Riemann tensor with pure, unabolishable by a coordinate transformation, gravitational waves."

Pirani presented a rough outline of this scheme at Chapel Hill in the discussion period following Bondi's presentation of Marder's work on gravitational waves. He employed Petrov's classification of the Riemann tensor into canonical types \cite{Petrov:1954aa}.\footnote{A translation into English of this article appeared in 2000 \cite{Petrov:2000aa}. Petrov  had classified such gravitational fields in his 1957 Moscow dissertation, but had already published his work in 1954. It would be appropriate here to cite Pirani's recollection of his encounter with Petrov's work, as recorded in an interview conducted by Dean Rickles \cite{Rickles:2011aa}. After arriving at the Dublin Institute for Advanced Studies in 1954, to work with Synge, he was asked to review the 1954 Petrov paper.  He recounts ` And I was reading the proofs of a book by Synge about special relativity where he said something about the geometrical characteristics of plane waves, and then boom! Connect Petrov, and it all comes out.' When asked if he managed to get a copy of the paper he responded `Yes, and I had somebody around here who helped to translate it, but he didn't do it very well.' } Pirani was motivated by a technical property of electromagnetic waves which as we shall see will play an important role later in the history of wave-like exact solutions of Eintein's equations.  They are represented by a null bi-vector  $F_{\alpha \beta}$ (the electromagnetic field) which has the property that it possesses only one eigenvector, and its  eigenvalue is zero, i.e., $F_{\alpha \beta} k^\beta = 0$, where $k^\beta$ is null and represents the propagating wave front. Could such a criterion be applied in the gravitational case? Petrov had exploited the bi-vector nature of the Riemann tensor, i.e., the anti-symmetry of $R_{\mu \nu \rho \sigma}$ under $\mu \leftrightarrow \nu$ and $\rho \leftrightarrow \sigma$, in addition to the symmetry under $\mu \nu \leftrightarrow \rho \sigma$  and the Bianchi identities to conceive of the Riemann tensor as a six dimensional symmetric matrix 
\beq
R_{AB} = \left(\begin{matrix}
     M &  N  \\
     N &  -M
\end{matrix} \right), \label{RAB}
\eeq
where  $N$ is a symmetric, traceless $3\times 3$ matrix.\footnote{ The index $A$ represents the pairs (23), (31),(12), (10), (20), (30).} $M$ is symmetric and in the vacuum case also traceless.
He then sought bi-vector eigenvectors of this matrix, and he discovered that the corresponding solutions of Einstein's equations split into three distinct classes, Types I, II, and III, corresponding respectively to six, four, and two independent eigen bi-vectors. The bi-vectors correspond to 2-surfaces, whose intersections yield 4-vectors which Petrov dubbed the Riemann principal vectors. The type III principal vector is null. Type II on the other hand has one null principal vector, with the others spacelike. Pirani noticed that the null surfaces spanned by the principal vectors were precisely those over which, according the the investigations of Lichnerowicz \cite{Lichnerowicz:1955aa} and Synge-O'Brien \cite{OBrien:1952aa}, discontinuities in the Riemann tensor could occur without violating Einstein's equations. Pirani therefore defined types II and III as representing gravitational radiation. His rationale, in part, was the fact that the freedom to alter these discontinuities corresponded to the similar freedom in electromagnetic theory to create waves that carried information.

But there is another related rationale, and it has to do with the energy content of a gravitational wave and its electromagnetic analogue. Electromagnetic waves are characterized by the fact that the flux of energy-momentum detected by observers moving at arbitrary speeds less than the speed of light is always non-zero. At first sight this appears not to be the case since the pseudo-tensor $t^\mu_\nu$ can be made to vanish along any arbitrary open curve in spacetime with a suitable change of coordinate system. Pirani's idea, which apparently came to him after he had written Schild,  was to employ his normal coordinates in a small neighborhood, and to calculate the average of $t^\mu_\nu$ over the surface of a small three-dimensional sphere enclosing this neighborhood. He showed that this average was an invariant under changes in the normal coordinates.  And in fact, for Petrov types II and III this average never vanished when viewed by local Minkowski observers moving along time-like geodesics. Such observers would always measure a non-zero flux of energy. It is curious that Pirani's assertions seem not to have elicited any discussion at the Chapel Hill meeting.

We conclude this section with some additional comments on the earlier history of the strong conservation law (\ref{strong}) that have implications for gravitational waves. This identity was in fact first derived by Einstein \cite{Einstein:1916ac}, based on the general coordinate invariance of the action. His claim that the pseudo-tensor $t^\mu_\nu$ could be interpreted as representing a local stress-energy content was almost immediately disputed.\footnote{See \cite{Kosmann-Schwarzbach:2011aa} for an account of this early history, with special attention to the criticism of Felix Klein \cite{Klein:1918aa}.} Bergmann included in his 1942 text \cite{Bergmann:1942aa} a derivation of the fact that the covariant divergence of the materiel stress-energy tensor density ${\cal T}_\mu{}^\nu$ could be written as an ordinary divergence,\footnote{The Lagrangians that have come into play to to this point are really scalar densities (up to a total divergence), and we should have represented them with the script notation. But to be consistent with the literature cited so far we have avoided using this notation. Here in discussing the strong conservation law it is crucial that we recognize we are dealing with densities. The relationship between the stress- energy tensor and the density is ${\cal T}_\mu{}^\nu =\sqrt{-g}T_\mu{}^\nu$}
$$
{\cal T}_\mu{}^\nu{}_{; \nu} = {\cal T}_\mu{}^\nu{}_{, \nu} - \Gamma^\rho_{\mu \nu} {\cal T}_\rho{}^\nu = {\cal T}_\mu{}^\nu{}_{, \nu} + \mathcal{t} _\mu{}^\nu{}_{, \nu},
$$
where $\mathcal{t} _\mu{}^\nu = 2\kappa \left(\delta_\mu{}^\nu {\cal G} - \frac{\partial {\cal G}}{\partial g^\alpha \beta}_{,\nu} g^{\alpha \beta}_{,\mu}\right)$ and ${\cal G}$ is the first order Einstein Lagrangian density,
\beq
{\cal G} = \frac{1}{2 \kappa}\sqrt{-g} g^{\mu \nu} \left(\Gamma^\sigma_{\rho \sigma} \Gamma^\rho_{\mu \nu} - \Gamma^\sigma_{\rho \mu} \Gamma^\rho_{\sigma \nu}  \right), \label{ELag}
\eeq
where $\kappa := 8 \pi G c^{-4}$.
 Thus ${\cal T}_{s\, \mu}{}^\nu{} := {\cal T}_\mu{}^\nu{}+\mathcal{t} _\mu{}^\nu$ is strongly conserved. Bergmann followed in 1949 with a general demonstration that in any generally covariant theory a strong conservation law could be derived from the Bianchi identity (\ref{Bianchi}) even when the field $y_A$ is coupled to matter characterized by a non-vanishing right hand side of the field equations. In this case the field equations are of the form ${\cal L}^A = {\cal P}^A$, and the Bianchi identities (\ref{Bianchi}) become\footnote{It is important to realize that this is still an identity, even though the field equations have been ``used". They have only been used in the sense that ${\cal L}^A$ does not vanish. So, for example, when we consider general relativity, we can think of the left hand side of Einstein's equations as simply defining a corresponding stress-energy tensor.}
$$
\left( F_{A \mu}{}^{B \nu} y_B {\cal P}^A \right)_{, \nu} + y_{A,\mu} {\cal P}^A \equiv 0.
$$
It follows using this identity and the general definition (\ref{pseudo}) of the pseudo-tensor (density) that
\beq
{\cal T}_{s \, \mu}{}^\nu:=  F_{A \mu}{}^{B \nu} y_B {\cal P}^A + \mathcal{t} _\mu{}^\nu, \label{strongtmunu}
\eeq
 has identically vanishing divergence, i.e.,
 \beq
 {\cal T}_{s \, \mu}{}^\nu{}_{,\nu} \equiv 0. \label{strongdiv}
 \eeq

We take the Einstein Lagrangian as an illustration. Considering an infinitesimal coordinate transformation $x'^\mu = x^\mu +\xi^\mu(x)$, we read off from $g'_{\mu \nu}(x) - g_{\mu \nu}(x)
=:\delta^* g_{\mu \nu} = - g_{\mu \nu , \alpha} \delta \xi^\alpha + 2 g_{\alpha (\mu} \delta \xi^\alpha_{, \nu)} =: - g_{\mu \nu , \alpha} \delta \xi^\alpha + F_{(\mu \nu) \rho}{}^{(\alpha \beta) \sigma} \delta \xi^\rho_{, \sigma}$ that
$ F_{(\mu \nu) \rho}{}^{(\alpha \beta) \sigma} = 4 \delta^\sigma_{(\mu} \delta^{(\beta}_{\nu)} \delta^{\alpha)}_\rho$. Therefore the first term in (\ref{strongt}) becomes $\kappa^{-1} \sqrt{-g} G_\mu{}^\nu= \sqrt{-g} T_\mu{}^\nu$, where $ G_\mu{}^\nu$
 is the Einstein tensor. The resulting strongly conserved stress-energy density is therefore
\beq
{\cal T}_{s \, \mu}{}^ \nu = {\cal T}_\mu{}^\nu + \mathcal{t} _\mu{}^\nu. \label{strongenergy}
\eeq

In 1951 Zatzkis \cite{Zatzkis:1951aa}, a student of Bergmann, was apparently the first to have shown in this context that as a consequence of the strong conservation law it was possible to introduce the superpotential employed above in (\ref{U}) by Goldberg.\footnote{ As already observed above, superpotentials had been employed by Freud \cite{Freud:1939aa}, as Zatkis noted, but not in direct relation to general covariance.} He  employed this superpotential in writing down the general relation between the time rate of change of the four-momentum within a volume and the flux of four-momentum through the enclosing two-surface. Zatzkis noted that because of the antisymmetry of the superpotential the time rate of change of the net four-momentum could be written as  a two-surface integral. The net four-momentum is
\beq
P_\mu := \int d^3 x {\cal T}_{s \, \mu}{}^0 =  \int d^3 x U_\mu{}^{[0 a]}{}_{, a} = \oint U_\mu{}^{[0 a]} n_a dS, \label{strongt}
\eeq
and it therefore follows from (\ref{strongdiv}) that the conservation law itself may be written as a sum of surface integrals,
\beq
\frac{dP_\mu}{dt} = \frac{d}{dt} \oint U_\mu{}^{[0 a]} n_a dS \equiv - \oint {\cal T}_{s \, \mu}{}^a n_a dS. \label{goldberg}
\eeq
Goldberg \cite{Goldberg:1953aa} and Zatzkis \cite{Zatzkis:1951aa} both cited Freud's expression for the superpotential,
\beq
{\cal U}_\mu^{[\nu \lambda]} = \frac{1}{2 \kappa} \sqrt{-g} g^{\sigma [ \rho} \delta^\nu_\mu g^{\lambda]\tau} g_{\rho \sigma, \tau}. \label{Ufreud}
\eeq
Goldberg noted in 1953 \cite{Goldberg:1953aa} the potential relevance of this relation for gravitational radiation. In 1955 \cite{Goldberg:1955aa} he observed that when one is outside a material source and the field equations are satisfied, then one can substitute the pseudo-tensor $\mathcal{t}_\mu{}^\nu$ for the strongly conserved ${\cal T}_{s \, \mu}{}^\nu$, and one obtains from (\ref{goldberg})
\beq
W_\mu := - \frac{dP_\mu}{dt} =  \oint \mathcal{t}_{ \mu}{}^a n_a dS, \label{mom}
\eeq
with the comment ``We shall say that gravitational radiation exists if the surface integral [(\ref{mom})] with $\mu = 0$ yields a finite result when the surface is infinitely far from the source points. The total flow of momentum may be similarly defined with $\mu = 1,2,3.$"\footnote{\cite{Goldberg:1955aa}, p. 1876}

\section{Trautman's dissertation work and follow up}

Almost immediately after Trautman began his tutelage under Infeld and Plebanski, he began publishing in rapid succession a series of papers in Infeld's newly established  {\it Bulletin de l'Academie Polonaise des Science}.\footnote{The first volume was published in 1954. Few Western institutions subscribed. Unfortunately, digitized versions of these early volumes are still not available. Thus it was not until Trautman visited London in 1958 that his work became more widely known in the West.} The first paper \cite{Trautman:1955ab} is a testament to his largely self-taught mathematical prowess, and also an early indication of interest in hyperbolic differential equations. He examined conditions which would insure the unique existence of solutions of a class of such equations in either the forward or backward domain of dependence, given data on a compact spatial hypersurface. These domains were bounded by characteristic null surfaces. A continuity equation played a critical role in his proof, and indeed related considerations would later come into play in his analysis of gravitational waves in general relativity. The second paper \cite{Trautman:1956ab} offers two extensions of the EIH method. The first is a proposal to include both even and odd powers of $\lambda$ in each $h_{ab}$ expansion. As we have seen, this idea had already been pursued by other authors. The second was a suggestion that arbitrary flat space curvilinear coordinates could be profitably employed in each step of the iterative expansion of $h_{ab}$. The advantage was impressively displayed in the next paper  \cite{Trautman:1956ac} in which Trautman used this extended EIH method to derive the exact $1/c$ expansion of the Schwarzschild metric in spherical coordinates, $g_{00} = 1 - \frac{2 k m}{c^2 r}$, and
 $g_{rr} = - \Sigma_{l=0}^\infty \frac{1}{c^{2l}} \left(\frac{2km}{r}\right)^l = - \left(1-  \frac{2km}{c^2 r} \right)^{-1}$.  Significantly, Trautman used a Dirac delta function representation of the stress-energy of the particle in this paper, a procedure that Infeld had finally adopted in 1954 \cite{Infeld:1954aa}.

 The next paper \cite{Trautman:1956aa} addresses a theme that Trautman will frequently revisit - the relationship between Lagrangian symmetries, conservation laws, and equations of motion. He follows up on Bergmann \cite{Bergmann:1949aa}, whom he cites, in investigating the implications on the equations of motion of the identities that follow from the scalar density nature of the relevant Lagrangian, namely,
\beq
\int \left( \frac{\partial {\cal L}_f}{\partial g^{\alpha \beta}} \delta^* g^{\alpha \beta} +  \frac{\partial {\cal L}_f}{\partial \psi_A} \delta^* \psi_A +  \frac{\partial {\cal L}_f}{\partial \psi_{A,\nu}} \delta^* \psi_{A,\nu} + \left( {\cal L}_f \delta \zeta^\nu  \right)_{, \nu}\right) d^4 x \equiv 0. \label{LagrangeIdentity}
\eeq
The Lagrangian field density ${\cal L}_f$ is assumed here to be that of a materiel field $\psi$.\footnote{We will use Trautman's later notion to distinguish the various different Lagrangians that come into play in his work.}
He assumes, as does Bergmann, that under the infinitesimal coordinate transformation $x'^\mu = x^\mu + \delta \zeta^\mu$, the `substantial' variation is\footnote{Trautman may be the first here to note in print that the `substantial' variation, defined as $\delta^* \psi_A(x):= \psi'_A(x) - \psi_A(x)$, is just minus the Lie derivative with respect to the field $\delta \zeta^\mu$. As Trautman pointed out in 2008 \cite{Trautman:2008ab}, the first general explicit definition of a Lie derivative was first written down by \'Slebodzi\'nski  in 1931 \cite{Slebodzinski:2010aa}. Also, as Trautman noted, it was van Danzig in 1932 who first assigned it the name 'Lie derivative', and Schouten used the term in his classic text, cited as a reference in the current article. The $\delta^*$ notion used by Trautman differs from Bergmann - and also Rosenfeld - who employed $\bar \delta$. Interestingly, this was also the notation employed by Noether in 1918 \cite{Noether:1918aa}, also cited by Trautman in the present article. Incidently, Rosenfeld originally employed this idea in 1930 \cite{Rosenfeld:1930aa,Rosenfeld:2017ab}. See \cite{Salisbury:2017aa} for additional historical background.}
 $$
 \delta^* \psi_A = - \psi_{A,\nu} \delta \zeta^\nu + F_{A \mu}{}^{B \nu} \psi_B \delta \zeta^\mu_{, \nu},
 $$
 where $F_{A \mu}{}^{B \nu}$ is constant.

Ultimately Trautman finds that if he has a field $\psi$, not gravitational, coupled to a perfect fluid in a curved spacetime, then the proper acceleration of a fluid element with proper four velocity $u^\alpha$, and stress-energy $\rho u^\alpha u^\beta $, can be written in terms of the variation of the field $\psi$ under general coordinate transformations.  Although Trautman does assume that the metric field is given, he does point out that if the dynamical gravitational action were added then the requirement that the covariant divergence of the net stress-energy vanish, $\left( T^{\alpha \beta} + \rho u^\alpha u^\beta \right)_{;\beta} = 0$ would follow as an integrability condition of the gravitational field equations. The symmetric form for the matter field stress-energy that Trautman employs is $T_{f\,\alpha \beta} := \frac{2}{\sqrt{-g}} \left(\frac{\delta {\cal L}_f}{\delta  g^{\alpha \beta}} \right) $.\footnote{ It was first written down independently by Belinfante \cite{Belinfante:1940ab} and by Rosenfeld \cite{Rosenfeld:1940aa}}. Using this condition Trautman finds ultimately that the equation of motion is
 \beq
 \rho u^\alpha_{;\beta} u^\beta = \left(g^{\alpha \nu} - u^\alpha u^\nu   \right) \left( \left(j^A \psi_B\right)_{;\beta} F_{A \nu}{}^{B \beta} + j^A \psi_{A;\nu} \right), \label{pem}
 \eeq
 where it is assumed that the field equations take the form $L_f^A := \frac{\partial L_f}{\partial \psi_A} - \left(\frac{\partial L_f}{\partial \psi_{A,\mu}}  \right)_{, \mu} = j^A$ and $L_f$ is the free $\psi$ field Lagrangian, coupled however to a given metric field, so that ${\cal L}_f = \sqrt{-g} L_f$. Trautman remarks in the introduction to this paper that his result is closely related to a recent paper by Infeld  \cite{Infeld:1955ab} in which imposition of the integrability condition can also be employed for a flat spacetime solution of linearized Einstein's equations to `derive' the equations of motion of a charged particle in interaction with an electromagnetic field. One other aspect of this paper that deserves note is that Trautman is able to obtain an antisymmetric object that in the flat case corresponds to Belinfante's \cite{Belinfante:1940ab} spin angular momentum density.  His deduction of the form of this object proceeds through the use of the identity (\ref{LagrangeIdentity})  in considering the contributions to the variation of the Lagrangian that follow by setting equal to zero the coefficient of $\delta \zeta^\mu_{,\nu}$.\footnote{For a discussion of the history of this technique, beginning with Felix Klein's assistant Bessel-Hagen \cite{Bessel-Hagen:1921aa}, see \cite{Salisbury:2017aa}.}

The next paper in the series, \cite{Trautman:1956ad}, extends the reasoning of the previous paper to spacetimes that possess Killing symmetries, but in addition to general coordinate symmetries also considers conformal and gauge symmetries. First he shows that a necessary and sufficient condition for conformal symmetry is the tracelessness of the stress-energy tensor, $T^\alpha{}_\alpha = 0$.  Turning to coordinate transformations, he shows that the variation of the material action corresponding to the fundamental identity (\ref{LagrangeIdentity}) may be integrated by parts to yield the ``strong" identity\footnote{This is the same reasoning employed by Bergmann \cite{Bergmann:1949aa} (and cited by Trautman) to obtain the ``generalized Bianchi identities". In the case of the gravitational action they are the contracted Bianchi identities.}
\beq
\left( T^\beta_f{}_\alpha  + L_f^A F_{A \alpha}{}^{B \beta} \psi_B \right)_{; \beta} + L_f^A \psi_{A;\alpha} \equiv 0. \label{ident2}
\eeq
Trautman then notes that if symmetries exist such that  $\delta \zeta^\alpha$ satisfy the Killing equation
$$
\delta^* g^{\alpha \beta}  = \delta \zeta^{\alpha; \beta} + \delta \zeta^{\beta; \alpha},
$$
(\ref{LagrangeIdentity}) yields up to ten conserved quantities when the field equations are satisfied through the resulting continuity equation
$$
\left(\frac{\partial {\cal L}_f}{\partial \psi_{A,\nu}} \delta^* \psi_A + {\cal L}_f \delta \zeta^\nu  \right)_{, \nu} = 0.
$$

Trautman's first paper dealing with gravitational radiation appeared in 1957 \cite{Trautman:1957ac}. Citing earlier work by Lichnerowicz \cite{Lichnerowicz:1955aa} and by O'Brien and Synge \cite{OBrien:1952aa}, he reviewed the discontinuities $\Delta$ in the metric that were permitted by Einstein's field equations.\footnote{Lichnerowicz indeed learned of these discontinuities from his thesis advisor Darmois \cite{Darmois:1927aa}.} They are permitted across characteristic null surface $\phi(x) = const$ with the property $\phi_{,\mu}\phi_{,\nu} g^{\mu \nu} = 0$.  Here he cites an as yet unpublished paper by  Pirani \cite{Pirani:1957aa} as having  taught him these results.\footnote{Trautman's paper was submitted for publication on November 24, 1956. He lists the Pirani paper as "in press" at {\it Acta Physica Polonica} although it was actually submitted to the {\it Physical Review} on October 18, 1956.} Trautman's main citation was an expression for the discontinuity of the Riemann tensor across $\phi$, which he showed to be of the form $\Delta R_{\alpha \beta \gamma \delta} = \frac{1}{2} \phi_{,[\alpha} h_{\beta][\gamma} \phi_{,\delta]}$ where he had previously shown that the discontinuity in the second derivative of the metric tensor was a genuine fourth rank tensor which took the form $\Delta g_{\alpha \beta, \gamma \delta} = h_{\alpha \beta} \phi_{, \gamma \delta}$, $h_{\alpha \beta}$ being itself a second rank tensor. Thus Trautman provided a fully covariant definition of the wave front of gravitational radiation.

In \cite{Trautman:1957aa} Trautman continues his exploration of the consequences of the assumption that the action is invariant under general coordinate transformations as in (\ref{LagrangeIdentity}), but he extends the action to include a singular particle contribution to the action of the form $W_p := \int d^4 x {\cal L}_p := \int d^4 x \int_{-\infty}^\infty \Lambda(\psi_A, \dot \xi^\alpha) \delta^4(x-\xi) ds$, where $\xi^\alpha(s) $ is the spacetime particle position parameterized by proper time $s$, and
$\dot \xi^\alpha := \frac{d \xi^\alpha}{ds}$. In so doing he is able to identify three different but related means for obtaining the particle equations of motion. The first uses a reparameterization invariant form of the particle action, $W_p = \int_{\lambda_1}^{\lambda_2} \Lambda(\psi_A,  \frac{d\xi^\alpha}{ds}) \sqrt{g_{\rho \sigma} \frac{d \xi^\rho}{d \lambda} \frac{d \xi^\sigma}{d \lambda}} d\lambda$ with parameter $\lambda$, and varies $\xi^\rho$.\footnote{Trautman appears to have been the first to exploit the use of reparameterization covariance in deriving the stress energy tensor of singular particles in general relativity, so we add some details that did not appear in this article. Defining $\xi'^\mu := \frac{d\xi^\mu}{d\lambda}$, we have $ds = \left(g_{\mu \nu} \xi'^\mu \xi'^\nu\right)^{1/2} d\lambda$. To obtain the stress energy tensor we need to calculate the variation of the action under the variation $\delta g_{\rho \sigma}$. Note first that $\delta \dot \xi^\mu = \xi'^\mu \delta  \left(g_{\alpha \beta}\xi'^\alpha \xi'^\beta\right)^{-1/2} = -\frac{1}{2}  \xi'^\mu  \left(g_{\alpha \beta}\xi'^\alpha \xi'^\beta\right)^{-3/2} \xi'^\rho \xi'^\sigma \delta g_{\rho \sigma} = -\frac{1}{2}  \dot \xi^\mu \dot \xi^\rho \dot \xi^\sigma \delta g_{\rho \sigma} $. Also, $\delta \left(g_{\alpha \beta}\xi'^\alpha \xi'^\beta\right)^{1/2} = \frac{1}{2}  \left(g_{\alpha \beta} \xi'^\alpha \xi'^\beta\right)^{1/2} \delta g_{\rho \sigma} \dot \xi^\rho \dot \xi^\sigma  $. So finally we find that $\delta W_p  = \frac{1}{2} \int d^4 x \int ds \left[- \frac{\partial \Lambda}{\partial \dot \xi^\nu}  \dot \xi^\nu    + \Lambda \right] \dot \xi^\rho \dot \xi^\sigma \delta g_{\rho \sigma}$, which delivers Trautman's expression for the resulting stress energy tensor.} 
Or one could vary the original particle action with the subsidiary condition $\dot \xi^\alpha \dot \xi_\alpha = 1$, or thirdly one could use the Einstein equation integrability condition $\left(T^{\alpha \beta}_f + T^{\alpha \beta}_p \right)_{; \beta} = 0$. He shows that all lead to the same particle equations of motion\footnote{Following up on the previous note, the variation of $\xi^\mu$ employs $\delta \Lambda = \frac{\partial \Lambda}{\partial \psi_A} \frac{\partial \psi_A}{\partial \xi^\sigma} \delta \xi^\sigma 
+ \frac{\partial \Lambda}{\partial \dot \xi^\mu} \left(g_{\alpha \beta}\xi'^\alpha \xi'^\beta\right)^{-1/2} \left(\delta^\mu_\sigma -g_{\rho \sigma} \dot \xi^\mu \dot \xi^\rho  \right) \delta \xi'^\sigma$ and $\delta \left(g_{\alpha \beta}  \xi'^\alpha \xi'^\beta\right)^{1/2} =  \left(g_{\alpha \beta}  \xi'^\alpha \xi'^\beta\right)^{-1/2} g_{\rho \sigma}  \xi'^\rho \delta \xi'^\sigma$ to deliver this result }
\beq
\Omega_\alpha := \frac{\partial}{\partial \xi^\beta}\left[ \left( \Lambda - \frac{\partial \Lambda}{\partial \dot \xi^\nu} \dot \xi^\nu \right)\dot \xi_\alpha   + \frac{\partial \Lambda}{\partial \dot \xi^\alpha}\right]\dot \xi^\beta - \frac{\partial \Lambda}{\partial \psi_A} \psi_{A;\alpha} = 0.  \label{eqm2}
\eeq

In the second part of the paper Trautman follows up on considerations of the previous one in examining conserved quantities which arise in the presence of Killing symmetries. In this case he effectively writes down the fundamental identity analogous to (\ref{LagrangeIdentity}), but now considering the sum of the two Lagrangian ${\cal L}_f + {\cal L}_p$,
$$
\frac{1}{2} {\cal T}_f^{\alpha \beta} \delta^* g_{\alpha \beta} + {\cal L}_f^A \delta^* \psi_A
+ \left(\frac{\partial {\cal L}_f}{\partial g_{\alpha \beta, \nu}} \delta^* g_{\alpha \beta}
+  \frac{\partial {\cal L}_f}{\partial \psi_{A, \nu}} \delta^* \psi_A
+ {\cal L}_f \delta \zeta^\nu \right)_{,\nu}
$$
\beq
+ \int_{-\infty}^\infty ds \delta^4\left(x-\xi  \right)\left[ -\Omega_\nu \delta \zeta^\nu +
\frac{d}{ds}\left\{ \left[  \left( \Lambda - \frac{\partial \Lambda}{\partial \dot \xi^\nu} \dot \xi^\nu \right)\dot \xi_\alpha   + \frac{\partial \Lambda}{\partial \dot \xi^\alpha} \right] \delta \zeta^\mu \right\} \right] \equiv 0. \label{ident3}
\eeq
Then assuming that $\zeta^\nu$ is an isometry Killing vector and assuming that both the field equations and the particle equation of motion (\ref{eqm2}) are satisfied, one can integrate over a spacelike hypersurface  $\sigma$ to get the conserved quantity
$$
\int_\sigma  \left(\frac{\partial {\cal L}_f}{\partial \psi_{A,\nu}} \delta^* \psi_A + {\cal L}_f \delta \xi^\nu  \right) d\sigma_\nu + \left.\left[  \left( \Lambda - \frac{\partial \Lambda}{\partial \dot \xi^\nu} \dot \xi^\nu \right)\dot \xi_\alpha   + \frac{\partial \Lambda}{\partial \dot \xi^\alpha} \right] \delta \zeta^\alpha\right|_\sigma = const.
$$
One reassuring application of this result is that if one simply has a single particle moving in a given gravitational field with no matter field, the function $\Lambda$ in this case is simply $\Lambda = - mc$, and the corresponding constants of the motion are simply $m \frac{d \xi_\nu}{ds} \delta \zeta^\nu = const $ while the particle equation of motion (\ref{eqm2}) gives a geodesic, $\frac{D \dot \xi^\alpha}{ds} = 0$. Trautman is likely the first to have obtained these results from the invariance of the matter field and particle action. It bears repeating, however, that it is assumed that the metric is given. There is no back reaction contemplated as yet.

The third part of the paper is a critique of Fock's claim that harmonic coordinates play a ``preferred" or ``privileged" role in general relativity. Trautman cites the 1955 Russian addition of Fock's classic text, issued first in English in 1959 \cite{Fock:1959aa}, in which Fock argues for the physical priority of the harmonic (or de Donder) coordinate conditions $\frac{\partial}{\partial x^\beta} \left(\sqrt{-g} g^{\alpha \beta}  \right) = 0$. Trautman argues that in the generic case in which there are no Killing symmetries this preference is not warranted. Infeld had already publically registered his own objections in Bern in 1955.\footnote{In discussion following Fock's talk Infeld says ``For three years my friend Professor Fock and I have discussed this problem, and we cannot convince one another. I doubt if we shall succeed in doing so today. But since Professor Pauli insists, I shall make a few remarks about the difference between Professor Fock and the usual understanding of relativity. Professor Fock adds to the gravitational equations the coordinate conditions ${\cal g}^{\mu \nu}{}_{,\nu} = 0$. In this way he obtains a system of equations which are invariant with respect to Lorentz transformations only. In this he sees a virtue, while I see in it a retrogressive step from the achievements of Relativity Theory." \cite{Fock:1956ab}, p. 240.  Infeld's view has been almost universally shared by the relativity community.}

In \cite{Trautman:1957ab} Trautman addresses for the first time the possible link between material motion and the production of gravitational waves. He asks whether it is possible that a compact gravitating system could emit periodic gravitational waves. His analysis is based on the strong conservation law (\ref{strongenergy}). He does not cite a source for this relation, but he does directly cite Goldberg 1955 \cite{Goldberg:1955aa} ( see (\ref{mom}) ) for the resulting ``power radiated at $x^0 = a$",
$$
W_0 = \oint_{S(a)} \left({\cal T}_{0}{}^k + \mathcal{t}_{ 0}{}^k  \right) n_k dS.
$$
The proof that the mean power radiated by a periodic, asymptotically Minkowskian gravitational field is zero follows almost immediately from this relation and the vanishing divergence (\ref{strongdiv}).  Assume that the temporal period is $\tau$, integrate the divergence over the infinite four-volume of thickness $\tau$ to find that
\beq
0 = -\int_{x^0 = a} \left({\cal T}_{0}{}^0 + \mathcal{t}_{ 0}{}^0  \right) d^3x + \int_{x^0 = a+\tau} \left({\cal T}_{0}{}^0 + \mathcal{t}_{ 0}{}^0  \right) d^3x
+ \int dx^0 \left({\cal T}_{0}{}^k + \mathcal{t}_{ 0}{}^k  \right) n_k dS = W_0. \label{mass}
\eeq
This leads Trautman to the conclusion, in response to the question posed by Pirani in \cite{Pirani:1957aa}, that gravitational radiation must be accompanied by mass loss of the source since as he observes the spatial integrals appearing in (\ref{mass}) ``for the Schwarzschild field ... are known to be proportional to the masses producing the fields."

 The next paper, \cite{Trautman:1958ad}, continues the investigation begun in \cite{Trautman:1957ac} of possible discontinuities of the Riemann tensor on characteristic (null) surfaces of Einstein's equations. Trautman writes down in generally covariant form the differential equation satisfied by this discontinuity on the surface. As an example of the resulting propagation of the discontinuity along a bicharacteristic Trautman derives the permitted discontinuity in a Schwarzschild geometry along the radial $r$ direction. Expressed with respect to orthonormal basis tetrads $e^\mu_\alpha$ the discontinuity satisfies
 $$
 \Delta R_{\alpha \beta \gamma \delta}(r) = \Delta R_{\alpha \beta \gamma \delta}(r_0) \frac{r_o - 2 m}{r - 2m}.
 $$
 Trautman notes that the $1/r$ behavior as $r \rightarrow \infty$ is ``characteristic for radiative phenomena."

 The next two papers, both written prior to Trautman's visit to London in May and June of 1958, propose proper boundary conditions near infinity to describe outgoing waves. The first \cite{Trautman:1958ab} focuses on scalar and electromagnetic radiation in Minkowski spacetime. In this case he chooses Sommerfeld's \cite{Sommerfeld:1949aa}  outgoing radiation condition which ``means that the system can lose energy in the form of radiation and that no waves are falling on the system from the exterior." Assuming the Poisson equation
 $$
 \left(\nabla^2  - \frac{\partial^2}{\partial t^2}\right) \phi = - 4 \pi \rho,
 $$
 for the scalar field $\phi$ with source $\rho$, the retarded solution takes the form
 $$
 \phi = r^{-1} \int_V \rho(\vec r', t - |\vec r - \vec r'|) dV' + {\cal O}(r^{-2}),
 $$
 and
 $$
 \phi_{, \sigma}  = k_\sigma r^{-1} \int_V \rho_{,0}(\vec r', t - |\vec r - \vec r'|) dV' + {\cal O}(r^{-2}),
 $$
 where $k^\sigma = (1, x^a/r)$. Thus Trautman proposes the boundary conditions $\phi = {\cal O}(r^{-1})$ and that there exists a $\psi = {\cal O}(r^{-1})$ such that $\phi_{, \nu} = k_\nu \psi +  {\cal O}(r^{-2})$. It follows, given that the scalar field stress-energy becomes
 $$
 4 \pi T_{\mu \nu} = \psi^2  k_\mu k_\nu + {\cal O}(r^{-3}),
 $$
 that the time rate at which energy-momentum radiated is
 $4 \pi \oint_S T_\mu{}^a n_a dS =  \oint_S \psi^2 k_\mu dS$.

 Similar outgoing radiation boundary conditions are formulated for electromagnetism. In this case Trautman shows that the appropriate conditions are $A^\mu = {\cal O}(r^{-1})$ and there exist four functions $B_\mu = {\cal O}(r^{-1})$ such that
 $A_{\rho,\sigma} = B_\rho k_\sigma +{\cal O}(r^{-2})$, and
 $B_\rho k^\rho = {\cal O}(r^{-2})$.

 The next \cite{Trautman:1958aa} and final paper before Trautman's visit to London in 1958 constitutes perhaps his most significant contribution in gravitational radiation theory. He writes down what he views as the analogue of Sommerfeld's outgoing radiation boundary conditions for gravitational waves, deduces a mass loss formula based on these assumptions, and confirms that the leading $1/r$ order of the Riemann tensor is type II. He assumes that there exist coordinate systems and functions $h_{\mu \nu} = {\cal O}(r^{-1})$ such that $g_{\mu \nu} = \eta_{\mu \nu} + {\cal O}(r^{-1})$, $g_{\mu \nu, \rho} = h_{\mu \nu} k_\rho + {\cal O}(r^{-2})$,  and $\left(h_{\mu \nu} - \frac{1}{2} \eta_{\mu \nu} \eta^{\rho \sigma} h_{\rho \sigma}  \right) k^\nu = {\cal O}(r^{-2})$. The null vector $k^\mu := n^\mu + t^\mu$ where $n^\mu$ is a unit outward point spacelike vector lying in a spacelike hypersurface $\sigma$ pointing perpendicular to the surface $r = const$ while $t^\mu$ is a unit timelike vector pointing perpendicular to $\sigma$.

 Trautman derived these results from the strongly conserved stress energy density (\ref{strongenergy}), citing also the Freud-Zatzkis-Goldberg superpotential (\ref{Ufreud}),
 \beq
 {\cal T}_{s \, \mu}{}^{ \nu} = {\cal T}_{\mu}{}^{ \nu}  + \mathcal{t} _{\mu}{}^{ \nu}
 \equiv \frac{1}{2 \kappa} \left(\sqrt{-g} g^{\sigma [ \rho} \delta^\nu_\mu g^{\lambda]\tau} g_{\rho \sigma, \tau} \right)_{,\lambda}. \label{strong2}
 \eeq
 But first he noted that if one were to employ the weaker boundary condition that had been advocated by Lichnerowicz  , namely $g_{\mu \nu,\rho} = {\cal O}(r^{-2})$, the total four-momentum $P_\mu$ calculated on a constant time slice (see (\ref{strongtmunu})),
 \beq
P_\mu = \int d^3 x\left( {\cal T}_\mu{}^0 + \mathcal{t} _\mu{}^0 \right) = \oint   {\cal U}_\mu^{[0 a]} n_a dS,
 \eeq
would be constant, independent of the time. To take into account the propagation of energy momentum he then considered a four-dimensional integration domain bounded by a timelike``cylinder" $\Sigma$ extended toward $r \rightarrow \infty$, and infinite spacelike hypersurfaces at times $x'^0 < x^0$, represented by $\sigma'$ and $\sigma$. Then employing the four-dimensional Gauss theorem one obtains from the vanishing divergence of (\ref{strong2})
\beq
P_\mu[\sigma] - P_\mu[\sigma'] = \int_\Sigma \mathcal{t} _\mu{}^\nu dS_\nu =: p_\mu,
\eeq
where it is assumed that the source does not extend to $\Sigma$, the timelike cylindrical boundary hypersurface at $r\rightarrow \infty$.  
Lastly, he looked at the asymptotic form of the curvature tensor. Assuming that the second derivatives are of the form $g_{\mu \nu, \rho \sigma} \approx 2 h_{\mu \nu, (\sigma} k_{\rho)} $ were of order $r^{-1}$, he deduced that there existed functions $ i_{\mu \nu} = {\cal O}(r^{-1})$ such that
there is therefore no contribution from ${\cal T}_{\mu \nu}$ on this surface. In order that this make sense, i.e., that the right hand side is finite, it was necessary for Trautman to show that the ${\cal O}(r^{-1})$ terms in the integrand canceled, as he did, using the third assumption on the behavior of $h_{\mu \nu}$. Furthermore, he was able to show that the resulting expressions for $P_\mu$ were invariant under coordinate transformations that preserved the boundary conditions, as is the power output $p_\mu$. In fact, he could show that $p_0$ was non-negative, proving that the radiating system loses mass.
$$
g_{\mu \nu, \rho \sigma} = i_{\mu \nu} k_\rho k_\sigma + {\cal O}(r^{-2}),
$$
obtaining the curvature tensor
\beq
R_{\mu \nu \rho \sigma} = \frac{1}{2} k_{[\mu} i_{\nu] [\rho} k_{\sigma]} + {\cal O}(r^{-2}). \label{R}
\eeq
It follows that $k_{[\mu} R_{\nu \rho] \sigma \tau} \approx 0$ and $ k^\mu R_{\mu \nu \rho \sigma} \approx 0$. Trautman notes that exact satisfaction of these conditions are those identified by Lichnerowicz \cite{Lichnerowicz:1958aa} and Pirani \cite{Pirani:1957aa} as characterizing pure gravitational radiation.

Pirani arranged for Trautman to present the content of these eleven papers at King's College London in May and June of 1958. He delivered a series of five lectures, the mimeographed notes of which were widely circulated.\footnote{The lectures were finally published in 2002 \cite{Trautman:2002aa}.} The ideas presented, and the resultant conversations with attendees - including Bondi, Pirani, and Ivor Robinson - strongly influenced research in gravitational radiation at King's College. The first lecture dealt with the boundary conditions in gravitational radiation theory, summarizing the contents of \cite{Trautman:1958aa,Trautman:1958ab}. The one notable addition is a reference to the exact gravitational plane wave solution of Petrov type II which had been recently rediscovered by Ivor Robinson, with the observation that this solution also was consistent with his own asymptotic results - as well as the characterization proposed by Lichnerowicz and Pirani.

Lectures two and three addressed the relation between equations of motion and gravitational radiation.  In addition to offering the same historical background and results reviewed in the series analyzed above, he addressed for the first time the issue of gravitational back reaction. These and new results were published following his visit to London \cite{Trautman:1958ac} . As had already been noted by Infeld \cite{Infeld:1938aa}, the gravitational fields of either the advanced or retarded wave solutions could be expanded in increasing powers of the radial distance from the source.  In fact, Infeld himself cited earlier electromagnetic expansions carried out by Nordstr\"om \cite{Nordstrom:1920aa} and Page \cite{Page:1924aa}. As had Infeld, Trautman observed that gravitational radiation appeared in the odd $k$ expansion terms ${}_kg_{00}$ and  ${}_kg_{ab}$, while in the even $k$ terms of  ${}_kg_{0a}$. Thus he continued with the `new approximation method' he had begun in \cite{Trautman:1956ab}. In the discussion of radiation and damping in his second and third lectures Trautman confined his attention to the EIH method in which surface integrals were performed around singularities, similar to the procedure discussed by Goldberg in 1955. He was very well aware of the fact that the metric solutions at each order  involved otherwise arbitrary functions of time that needed to be matched with the presumed asymptotic behavior of retarded gravitational waves. I will postpone a discussion of this matching for later.  In the later 1958 paper he took a different tact in working with the particle contributions to the Lagrangian that had been promoted by infeld. I will also discuss this approach later. It turns out that Trautman's results regarding the damping term in the equation of motion of two equal masses was incorrect by a factor of two.
  
 Lecture four dealt with the propagation of gravitational disturbances (reviewing the analysis of discontinuities of \cite{Trautman:1957ac}, conservation laws and symmetry (referring to \cite{Trautman:1957aa}), and the `fast' motion approximation method for investigating gravitational radiation.

\section{Robinson, Trautman, and gravitational wave solutions}

The period following Trautman's visit to London in 1958  witnessed enormous progress in techniques for finding wavelike solutions of Einstein's equations. Contributors were scattered over a wide geographical range, but with growing international contact and collaboration. Ivor Robinson and Andrzej Trautman were among the central players, and  we will attempt in this section to place their role in these developments in this wider context.

Ivor Robinson had no formal training in general relativity, and by his own account he privately studied Peter Bergmann's text \cite{Bergmann:1942aa}, where Bergmann's remarks concerning gravitational plane waves led to discussions with Bonnor and Pirani.\footnote{\cite{Rindler:1987ab}, p. 14. According to interviews of Robinson conducted by Richard Matzner in March, 2014 \cite{Matzner:2019aa}, Bonnor, based at the University of Liverpool, was an important early influence. Regarding Bergmann, he concluded that the active areas of research were the study of equations of motion - which he found to be `elegant but too messy' - and unified field theory `which appealed to him'. This would be consistent with his readiness later on to look for implications of his research in Einstein-Maxwell theory. } Already in 1954 he began to investigate what appeared to him to be a related question: what would be the gravitational field of a photon?\footnote{\cite{Rindler:1987ab}, p. 10 }
This question led to his key rediscovery in early 1956 of exact plane wave solutions of Einstein's equation. He was not in fact at the time aware that Brinkmann had already published these solutions in 1925 \cite{Brinkmann:1925aa}. Robinson later published an account of his steps leading to these solutions \cite{Robinson:1985ab}.\footnote{See also the story presented by Rindler and Trautman in 1987 \cite{Rindler:1987ab}.} Due to the appearance of a singular metric in their planar vacuum solutions of the field equations, Einstein and Rosen had argued that exact plane wave solutions did not exist \cite{Einstein:1937aa}. Rosen's metric as cited by Bondi \cite{Bondi:1957aa} takes the form\footnote{Rosen actually does not give this explicit expression in \cite{Rosen:1937aa} although it can be deduced from his analysis. }
$$
ds^2 = e^{2 \phi} \left(dt^2 - dz^2 \right) - u^2 \left(e^{2 \beta} dx^2 + e^{- 2 \beta} dy^2 \right),
$$
where $u = t - z$, $ \beta = \beta(u)$ and $\phi = \phi(u)$. The metric is singular when $u \rightarrow 0$.\footnote{There is a fascinating story behind the Einstein Rosen and Rosen papers, revealed through the research of Kennefick \cite{Kennefick:2007aa}. He found that the original draft of the Einstein-Rosen paper - which as far as  we can tell is no longer extant - had been submitted to the Physical Review. The referee H. P. Robertson had recommended revisions which Einstein rejected. Yet the paper eventually published in the Journal of the Franklin Institute did incorporate Robertson's suggestion (made later orally to Einstein) that this metric could be reinterpreted as representing a cylindrically symmetric spacetime, in which case $u$ would represent a radial coordinate.}

Returning to Robinson,  his attachment to null geodesics would continue throughout his career. Their role in describing the symmetry of plane electromagnetic wave solutions in flat spacetime suggested to him in 1954 an approach to finding the gravitational field of a photon. He knew that a plane wave traveling in the $z$ direction was invariant under the group of Lorentz transformations including elements that left the null $u:= t-z = const$   surfaces fixed (null rotations) and displacements along the null direction. The null rotations are of the form
$$
u' = u,
$$
$$
v' = v +  \xi x +\eta y +\frac{1}{2}\left( \xi^2 + \eta^2 \right)u,
$$ 
$$
x' = x + \xi u,
$$
\beq
y' = y + \eta u \label{nullrot}
\eeq 
with arbitrary parameters $\xi$ and $\eta$. As shown by Synge \cite{Synge:1956ab}, representing the null direction by $k_\mu$, with $k_\mu k^\mu = 0$, the plane wave was said to be null when the field tensor $F_{\mu \nu}$ and its dual $F^{* \mu \nu} := \frac{1}{2}\epsilon^{\mu \nu \rho \sigma} F_{\rho \sigma}$ satisfy $F_{\mu \nu} F^{\mu \nu} = F_{\mu \nu} F^{*\mu \nu} = 0$.  Then it takes the form
$$
F_{\mu \nu} = B(u) \left[\left(k_\mu l_\nu - k_\nu l_\mu \right) \cos \theta(u) + \left(k_\mu m_\nu - k_\nu m_\mu \right) \sin \theta(u)\right],
$$
where $l_\mu$ and $m_\mu$ are orthogonal to each other and to $k_\mu$. This expression further involves bivectors, antisymmetric second-rank tensors that will also play a fundamental role in Robinson's career.
Robinson's idea was to find the general form of a spacetime metric that satisfied a similar symmetry. His starting point was to perform  the active symmetry transformations (\ref{nullrot}) on the line $u = t - z =: \rho$ with $x = y = v = 0$, thereby covering Minkowski space with new coordinates defined in part through null rotations, yielding $x' =  \xi \rho$, $y' = \eta y$, $v' = t' + z' =  \frac{1}{2} \left(\xi^2 + \eta^2\right) u$.\footnote{The nature of these rotations was apparently not known to Synge. According to Trautman, \cite{Trautman:2009aa}, p. 1200, Robinson pointed out to Pirani that Synge had mistakenly maintained in the first edition of his special relativity textbook that all proper Lorentz transformations could be represented as a product of a boost and an orthogonal spatial rotation - a `4-screw'. Null rotations are not 4-screws.} In addition he translated $v'$ by a $\sigma$, yielding $v' = \frac{1}{2} \left(\xi^2 + \eta^2\right) u+ \sigma$. The inverse transformations are $\xi = x/u$, $\eta = y/u$, $\rho = u$, and $\sigma = v - \frac{x^2 + y^2}{2 u}$ and the metric in these new coordinates is
$$
ds^2 = 2 d\rho d\sigma - \rho^2 \left(d\xi^2 + d\eta^2  \right).
$$
Of crucial import here is the observation that the coordinate transformation is singular at $\rho = 0$ and that the singular line element represents a coordinate breakdown. Next he considered the most general spacetime line element that was invariant under the analogous transformations $\left(\xi,\eta,\rho,\sigma\right) \rightarrow \left( \xi + Y^1, \eta + Y^1, \rho, \sigma + Z \right)$ and $\xi \rightarrow \xi \cos \Theta - \eta \sin \Theta$, $\eta \rightarrow \xi \sin \Theta + \eta \cos \Theta$. It took the form $ds^2 = A(\rho) d \rho^2 + 2 B(\rho) d\rho d\sigma + C(\rho) d\sigma^2 - R(\rho)^2 \left(d\xi^2 + d \eta^2 \right)$. After some false starts he finally settled on a less restrictive Einstein-Maxwell  line element of the form $ds^2 = 2 e^{\phi(\rho)} d\rho d \sigma - \rho^2 \left( d\xi^2 + d\eta^2  \right)$. This is more general in the sense that  the Ricci tensor $R_{\mu \nu}$ defines a source stress-energy tensor which one can interpreted as electromagnetic with the identification of the electromagnetic potential as $A_{,\mu} = \rho \left[\xi \alpha(\rho) + \eta \beta(\rho)\right]\rho_{,\mu}$ with $\frac{d\phi}{d \rho} = \frac{\rho}{2} \left(\alpha^2 + \beta^2\right)$. 
And it was here that, in his words, he ``finally realized exactly what I did have in the [electrovac] solutions: a non-singular plane wave. The way I had set things up, there was indeed a singularity of the prescribed kind at $\rho = 0$; but one could choose the functions $\alpha$, $\beta$, and $\phi$ so as to make the wave zero and the metric Minkowskian in the neighborhood of the singularity. I wondered if the trick would work for a purely gravitational wave. So I tried the line element
$$
2 e^{\phi(\rho)} d \rho d \sigma - \rho^2 \left[e^{2 \theta(\rho)} d \xi^2 + e^{-2 \theta(\rho)} d \eta^2\right].
$$
Here the field equations reduced to $\frac{d\phi}{d\rho}= \rho \left(\frac{d\theta}{d\rho}\right)^2$. Again we have a typical Rosen singularity at $\rho = 0$, and again we can choose the functions $\theta$ and $\phi$ so that the metric is Minkowskian in the neighborhood of the singularity."\footnote{\cite{Robinson:1985ab}, p. 416} The most general line element that included both gravitational and electromagnetic plane waves was of the form
$$
ds^2 = 2 H(\xi, \eta, \sigma) d\sigma^2 + 2 d \rho d\sigma - d\xi^2 - d\eta^2.
$$

(A natural alteration of this metric apparently led to Robinson's first publication - in Infeld's journal - after he reported on his discovery, in the Spring of 1959 in Warsaw,  of a spacetime with constant electric and magnetic fields \cite{Robinson:1959aa}.)

W. B. Bonnor, then at the University of Liverpool, was present for Robinson's 1956 lecture, and he reported on Robinson's discovery in 1957 \cite{Bonnor:1957ab}. Robinson never did publish this work under his exclusive name, but was finally persuaded to join Bondi and Pirani in a joint publication on plane waves in 1959 \cite{Bondi:1959aa}. Bondi had in the meantime independently found the plane waves and reported his results in 1957 \cite{Bondi:1957aa}. The invariance under a five-parameter group modeled on the flat electromagnetic case features in this discussion.

As background for the remainder of this tale we  list here the affiliations of several other relativists whose interactions with Trautman and Robinson will feature in this brief overview of advances in gravitational wave physics in the late 1950's and early 1960's. We are already acquainted with Pirani  who obtained his second Ph. D. with Bondi and Hoyle at Trinity College, Cambridge, in 1956 - after spending 1954-55 with Synge at the Dublin Institute for Advanced Studies.  He lectured at King's College from 1955 to 1960, but with research stays at the Advanced Research Laboratory of Martin-Marietta Corporation (RIAS) in Baltimore in 1958 and at the University of North Carolina in 1958-59. His long-term appointment at King's College began in 1960. Roy Kerr was a student at Cambridge from 1955, receiving his Ph. D. in 1958. He regularly attended seminars at King's College.\footnote{See \cite{Kerr:2009aa}} It was there that he met Peter Bergmann who was visiting in 1958.\footnote{Salisbury interview with Kerr, Dec. 13, 2013}  Bergmann offered him a research appointment at Syracuse where he remained for one year before joining Goldberg at Wright Patterson from 1960 to 1962. He was a visitor in Austin beginning in 1962 prior to becoming a regular faculty member from 1964 to 1969. Roger Penrose will of course also feature in this discussion. He completed his degree in Mathematics at Cambridge in 1957 with John Todd, although his unofficial `advisor' was Dennis Sciama.\footnote{Interview Salisbury and Rickles conducted with Penrose June 16, 2017. Penrose's father was a friend of Sciama, and Roger was first introduced to him at a lunch where Felix Pirani was also present.} From 1957 to 1959 he held a fellowship at St. John's College, followed by post-doctoral NATO research fellowships at Princeton University, 1959-1960, and Syracuse 1960-1961. He returned to King's College as a research associate from 1961 to 1963, whence he was appointed 1963-64 to a faculty position in Austin.

Next  we turn to Pascual Jordan  at the University of Hamburg, Germany. He oversaw a group that had initially been concerned with cosmology. Engelbert Sch\"ucking began studying with him in 1952, completing a Ph. D. under his direction in 1956.\footnote{Salisbury interview with Sch\"ucking, July 12, 2006. } For the next two years he worked with Otto Heckmann at the Hamburg Observatory, before obtaining a  post-doctoral appointment at Syracuse University in the Spring of 1961.\footnote{Sch\"ucking interview.} J\"urgen Ehlers began his studies with Jordan in 1954, obtaining his Ph. D. in 1957. He actually accompanied Jordan to the 1955 Bern meeting where he was impressed with Bergmann's authoritative overview of progress in quantum gravity.\footnote{ Salisbury interview with Ehlers, March 19, 2008.} In 1958 he was sent by Jordan to visit Bergmann in New York to make arrangements for an interchange of young relativists that would be financed by the German industrialist Friedrich Flick. During his stay he attended a Stevens Institute meeting and was asked by Bergmann to report on the work of the Jordan group in Hamburg.\footnote{Ehlers interview. } In a subsequent short visit to Syracuse he talked briefly to Rainer Sachs about the possibility of Sachs going to Hamburg for a post-doctoral research position.\footnote{Salisbury private communication with Sachs.} Ehlers remained in Hamburg until 1962 when he returned to Syracuse on a Flick Fellowship during the upcoming academic year. In 1958 Rainer Sachs was close to completion of his degree under Bergmann's direction. In fact, he went to Hamburg already in late 1958 with a one year Flick Fellowship and Ehlers was his `de facto supervisor', \footnote{Private communication with Sachs.} finally receiving his Ph. D. in 1959 after returning to Syracuse. From 1960 to 1961 he was a post-doctoral fellow at King's College, working with Felix Pirani. Wolfgang Kundt started working with Jordan at about the same time as Sch\"ucking, completing his Ph. D. in 1959. He himself went to Syracuse in January 1959 on a Flick Fellowship, returning to Hamburg in Spring, 1960. Istvan Ozsvath, also a member of the Jordan group, received his Ph. D. with Heckmann in 1960. He joined Alfred Schild at the University of Texas at Austin in 1962. In this same year, the fifth member of the group, Manfred Tr\"umper,   completed his Ph. D. under Jordan. There followed a post-doctoral appointment at Syracuse in 1962-63. As we shall see, the central focus of the Jordan group was the development of a strategy for constructing exact solutions of Einstein's equations.

There are also Belgian, French and Spanish connections. Jules G\'eh\'eniau and Robert Debever both received their Ph. D.'s at the University of Brussels with Th\'eophile de Donder, himself a student of Henri Poincar\'e. Michel Cahen received his doctorate at Brussels in 1960 under the tutelage of both Lichnerowicz and Debever. He was a Frick Fellow in Hamburg in 1961. The  Catalonian Llu\'s Bel  received his Ph. D. with Lichnerowicz in Paris in 1960 and was a research assistant at the University of North Carolina from 1961-62 and a visiting lecturer at the University of Texas at Austin 1962-63.

We have already encountered Achilles Papapetrou in regard to particle equations of motion. He perhaps more than anyone else personifies the international character of relativity research in the late 1950's.\footnote{See \cite{Hoffmann:2017aa}} Born and educated in Greece, he held positions at the Dublin Institute for Advanced Study 1946-48 and the University of Manchester, England, 1948-1952, before joining the Research Institute for Mathematical Physics of the (East) German Academy of Science in Berlin in 1952. Before leaving Berlin in January 1962 to become the Director of Research at the Centre Nationale de la Recherche Scientifique (CNRS) in Paris he had profited from several research stays in Paris and London. His major doctoral student in Berlin was Hans-J\"urgen Treder who had actually completed his thesis in 1954 though the degree was not awarded until 1956.

We return now to the Bondi, Pirani, and Robinson article.  It was submitted for publication in October, 1958, after Trautman's visit to London. The route to the plane wave metric was more direct than that described above by Robinson. The authors proceeded with the expectation that the Weyl tensor they sought should exhibit an invariance under a five-dimensional group similar to that of the null electromagnetic field in flat spacetime.  They simply appealed to Petrov \cite{Petrov:1954aa}, citing as the relevant metric
$$
ds^2 = e^{2 \phi} \left(d\tau^2 - d \xi^2 \right) - u^2 \left[\cosh (2 \beta) (d\eta^2 + d\zeta^2) +\sinh (2 \beta) \cos (2 \theta) (d\eta^2 - d \zeta^2)\right.
$$
$$
- \left.2 \sinh (2 \beta) \sin (2 \theta)  d \eta d\zeta \right],
$$
where $\phi$, $\beta$ and $\theta$ are functions of $u:= \tau - \xi$ 
No derivation was offered, and it is unlikely that Robinson himself would have deduced this form from the Petrov type. He always placed more emphasis on the behavior of null congruences, employing a method that was first spelled out in detail in published form in the contribution that Robinson and Trautman made to the 1962 Warsaw meeting. However, one finds reference to it in particular in the work of Wolfgang Kundt, reported first in 1960 in his joint publication with Jordan and Ehlers \cite{Jordan:1960aa}.\footnote{An English translation of this article was published as a Golden Oldie in 2009 \cite{Jordan:2009aa} with commentary by George Ellis \cite{Ellis:2009aa} .}  This was the first article in a series from the Hamburg group on exact solutions of Einstein's equations that has become known as the Hamburg Bible. But it appears that this term originally referred to a mimeographed one hundred twenty-five English summary of the work of the Hamburg group that included only this first paper of the series. We make this judgement based on the copy in Robinson's files that carries this title. It appears to have been distributed in 1960. In the introduction Jordan refers to previous work by himself, Ehlers, Kundt, and Sch\"ucking. Sch\"ucking comes into play in conjunction with work he did with Heckmann in giving a description of relativistic hydrodynamics \cite{Schucking:1957aa}\cite{Schucking:1959aa}. In doing so they built on the work of Lichnerowicz and Synge.  Section II, probably written by Ehlers, expands on his own 1957 dissertation work with relativistic hydrodynamics, defining the properties of expansion, rotation, and vorticity for ideal fluids moving along time-like congruences and deriving consequences for exact cosmological solutions. This copy actually predates Ehlers own published work on this topic which appeared in the 1962 Infeld Festschrift \cite{Ehlers:1962ac}. This is especially relevant to our story since, as noted above, Sachs began to work under Ehlers's tutelege at the end of 1958.  Already in 1959 they looked at the null geodesic congruence associated with null electromagnetic fields \cite{Ehlers:1959aa}. This was the beginning of their study of the geometry of null congruences \cite{Ehlers:1962aa} that we will revisit shortly.

First we return to the Jordan Ehlers Kundt article. It is significant on the one hand for its comprehensive treatment of the Petrov classification. Petrov had himself, already in 1954, reduced the problem to the determination of eigenvectors of a complex $3 \times 3$  matrix $P := M + i N$ (see equation \ref{RAB}). The resulting eigenvectors with their respective eigenvalues come in complex conjugate pairs. Petrov's classification was refined by Bel in \cite{Bel:1958ab} and \cite{Bel:1959aa}.\footnote{Bel's comprehensive overview of his work \cite{Bel:1962aa} given as a series of lectures at the Coll\`ege de France in December, 1960, has been published in English translation as a Golden Oldie \cite{Bel:2000aa}. See \cite{Senovilla:2000aa} for an editorial note with a short Bel biography.} As did Petrov, Bel worked with the Weyl tensor in an orthonormal frame. In seeking the eigenvectors of $P$ he repeated Petrov's analysis in distinguishing three cases where the complex eigenvalues always sum to zero. In case I they are distinct. In case II two of them coincide, with the third equal to minus twice this common value. Case III has all three eigenvalues vanishing. But Bel refined this scheme further in devoting one of his Coll\`ege de France lectures to a critique of Pirani's classification. Pirani himself utilized the complete scheme in his Infeld Festschrift review article that we shall address shortly. Type II generally has the canonical form
\beq
P = \left(\begin{matrix}
     -2m - 2n i& 0 & 0  \\
     0 &  m + \sigma + n i   & \sigma i \\
     0 & \sigma i & m - \sigma + n i
\end{matrix} \right), \label{P}
\eeq
but Bel identified as type IIa, now known as type N, the special case where $m = n = 0$. In this case the one independent complex eigenvalue vanishes. Pirani acknowledged the extensive contributions of Debever and Geh\'eniau \cite{geheniau:1956ab}\cite{Geheniau:1956ac}\cite{Debever:1958aa} in the development of this theory.

Jordan, Ehlers and Kundt  laid out systematically this Petrov classification, observing that the original complex Petrov approach actually constituted a representation of  the anti-self dual Weyl tensor, constructed by adding to the Weyl tensor $i$ times its dual. The eigen bivectors are themselves self-dual.\footnote{See for example the comprehensive treatment in \cite{Stephani:2003aa} .} The self-dual notion first arose in relation to null electromagnetic fields where one could form the complex sum of the null field  tensor $F_{\mu \nu}$ with $i$ times its dual $\tilde F_{\mu \nu}$ to get a complex bivector whose dual is $-i$ times the original sum. This construction could then be readily extended to both the left and right duals of the Weyl tensor. There is a natural correspondence between these objects and spinors, and in fact the Petrov classification of the Weyl tensor takes a much simpler form in this language as was shown in 1960 by Penrose \cite{Penrose:1960aa}. It is a curious fact that Robinson had already found complex self-dual bivectors useful before the advent of spinor theory - and according to Wolfgang Rindler never really embraced their use.\footnote{ Salisbury private communication. There is, however, a lingering puzzle. There exists in Robinson's papers a draft proposal, written most likely in 1963, by Rindler, Robinson, and Oszvath which is entitled Research in the Application of Spinors and Other Representations of the Lorentz Group to General Relativity. We  do not know if the proposal was submitted and or funded. Rindler had of course at that time already made substantial progress with the collaboration with Penrose that eventually resulted in their magisterial books on the subject\cite{Penrose:1986aa}\cite{Penrose:1988aa} .}  But perhaps the most significant contribution of this paper was the discussion by Kundt of the Robinson plane wave solution. It is prefaced by the remark `The following deduction is due to I. Robinson. One of the authors (K.) wants to express his thanks for oral communication.'\footnote{``Der folgende Herleitungsgang stammt von I. Robinson. Einer der Verfasser (K.) m\"ochte seinen Dank f\"ur die m\"undliche Mitteilung aussprechen."} The authors had proven that the Weyl tensor of a type N field assumed the form $R_{hijk} = m_{hi} m_{hi} - \tilde m_{hi} \tilde m_{hi}$ where $ \tilde m_{hi}$ is the dual of $m_{hi}$. The bivector $m_{hi} = l_{[h} \xi_{i]}$ where $l_h$ is null and covariantly constant, $l_h l^h = 0$ and $l_{h;i} = 0$. Also it is orthogonal to the spatial vector $\xi_i$, $l_i \xi^i = 0$. It follows that $m^{jk}$ is an eigen vector of $R_{hijk}$ with eigenvalue zero, $R_{hijk} m^{jk} = 0$. Robinson deduced that, in an appropriate gauge, a covariantly constant $W^{hi} = l^{[h} \xi^{i]}$  and $\tilde W^{hi} = l^{[h} \eta^{i]}$ could be constructed using $ m_{hi}$ and its dual. Then for a suitable choice of coordinates the  triads  $l^h$, $\xi^h$, $\eta^h$ could be chosen as basis vectors for a coordinate system. A necessary and sufficient condition for this was that the corresponding infinitesimal commutator vanished, and this was demonstrated. From there Robinson deduced that the line element took the form
$$
ds^2 = 2 du dv + dy^2 + dz^2 + 2 H(u,y,z) du^2.
$$
This paper was received by the Mainz journal in October, 1959. Kundt was still at Syracuse at that time, but he did meet Robinson at the colloquium on Les Th\'eories Relativistes de la Gravitation that took place at Royaumont Abbey in France, June 21 - 27, 1959 \cite{Lichnerowicz:1962ab}.  We assume that Robinson spoke on this topic there, but we  have only indirect evidence.\footnote{Kundt has confirmed to Salisbury in a private communication that it was at Royaumont that he learned of this construction procedure from Robinson.} Unfortunately he never submitted his presentation for publication, and we are in possession of neither an abstract nor a title. However, Trautman did present this line element in his own Royaumont paper \cite{Trautman:1962ad}, attributing it to Robinson in a paper that was `en cours de publication'.

Royaumont was a significant meeting.  With the exception of Jordan and Cahen, all of the relativists mentioned above took part. Its significance is evident in an unpublished report in English by Jordan, Ehlers, and Sachs, undated but apparently from 1960. The second section by Sachs on gravitational radiation refers to reports at the meeting by Lichnerowicz, Bel, Penrose, and Robinson.\footnote{The report, entitled `Progress in the field of general relativity' is in Robinson's papers. Evidence for its date is gathered from its reference to the Infeld volume whose date is given as 1960. Also Sachs's section on gravitational radiation is a preview of \cite{Jordan:1961aa} which is `in preparation.' This paper was actually submitted in October, 1960. } Robinson had already made contact with several  on the occasion of his visit with Trautman in Warsaw in the Spring of 1959, most notably Ehlers and Sch\"ucking in Hamburg.

We suspect that it was as a result of conversations at Royaumont that Pirani and Robinson agreed that they would write together a review of the then current status of gravitational radiation theory for inclusion into the Infeld Festschrift. The article did appear, but with the sole authorship of Pirani \cite{Pirani:1962ad}. In his acknowledgements appears the statement `The article suffers from the absence, as co-author, Mr. Ivor Robinson, who was prevented by circumstances beyond his control from collaborating, and to whom the writer is indebted in all questions concerning the theory of gravitation to an extent which is impossible to acknowledge.' A report by Pirani \cite{Pirani:1960aa} , dated August, 1960, concludes with the note: `For further information the reader is referred to the forthcoming article by the author and Robinson in the Warsaw presentation volume.' A earlier draft with joint authorship  exists in Robinson's papers - in addition to four pages entitled `Corrections and revisions (Trautman and Sachs)'. We  do not know what were the circumstances beyond his control which prevented Robinson from collaborating.\footnote{Sachs has offered the observation to Salisbury in private communication that he was not aware of any tension between Robinson and Pirani.}  One notable addition to the paper in regard to null geodesics, at Sachs's suggestion, was that it should give more credit to the Hamburg group and to Bel and Lichnerowicz. Related to this suggestion is a condition identified in the final version as the `Robinson equation'. Pirani's remark in his note to Robinson says of this equation,
$$
k_{(a;b)} k^{a:b} = \frac{1}{2} \left(k^a{}_{;a}\right)^2,
$$
where $k_a$ is the propagation vector,
`that it could be referred to `as the ray-divergence equation, or something?'. This is now known as the condition for vanishing shear and it seems likely that it was to have first appeared in print in Robinson's Royaumont lecture. Pirani refers both to this lecture and to Robinson's 1956 King's College lecture as a source.

Jordan, Ehlers, and Sachs (JES) published in 1961 \cite{Jordan:1961aa} the foundational paper on the geometric properties of null congruences.\footnote{A translation into English has appeared as a Golden Oldie \cite{Jordan:2013aa}, with editorial commentary by MacCallum and Kundt \cite{MacCallum:2013aa}. } The work was submitted for publication in October, 1960. The authors carefully defined the so-called optical scalars of divergence, twist, and shear, and gave a clear interpretation of them in terms of the shadow cast by a small opaque object. In attempting to  determine precedence in terms of submission dates, as far as  we can tell the first explicit expression for the shear of a null congruence appeared in Robinson's 1961 article \cite{Robinson:1961aa} which was received in September, 1960. The JES article was submitted in October, 1960. JES provide additional evidence of Robinson's familiarity with shear, stating that the shear-free property of both the null congruence associated with null electromagnetic fields \footnote{footnote 43 in \cite{Jordan:2013aa} and footnote 2 on p. 37 in \cite{Jordan:1961aa}  } and N type waves\footnote{footnote 45 in \cite{Jordan:2013aa} and footnote 1 on p. 38 in \cite{Jordan:1961aa}} was demonstrated by Robinson. The references are to his Royaumont lecture. Robinson proved in  \cite{Robinson:1961aa} that a geodesic null  congruences is shear-free if and only if the associated bivector field is a solution of Maxwell's equations for charge-free space.  He showed that the congruence represented by $k_i$ is geodesic if and only if there exists a complex scalar field $\xi$ satisfying
\beq
\left\{ k_{[k;l]} - i {}^*k_{[k;l]} \right\} k^l = \left(\frac{1}{2} k^l{}_{;l} + \xi \right) k_k, \label{kkl}
\eeq
where ${}^*$ signifiies the dual. He follows this relation with the remark\footnote{\cite{Robinson:1961aa}, p. 291} ``As Dr. R. K. Sachs has in effect shown, the rates of rotation, dilatation, and maximum shear are proportional to $i (\xi - \bar \xi)$,  $(\xi + \bar \xi)$, and $\gamma$ where
\beq
\gamma = \left[ 2 k^{p:q} k_{(p;q)} - \left(k^i{}_{;l} +  \xi + \bar \xi\right)^2 - \left(\xi + \bar \xi\right)^2\right]^{1/2}".
\eeq
JES made extensive use of an orthonormal tetrad consisting of the null $k_i$, another null vector $n_i$ and a complex spatial vector $m_i$, satisfying the relations $k_i n^i = 1$, $m_i \bar m^i = -1$ and all the remaining scalar products vanishing. They were motivated by the fact that these vectors bear a simple relation to spinors. Contracting (\ref{kkl}) with these vectors delivers the optical scalars in their more familiar form, $\omega = i (\xi - \bar \xi) = \left[ \frac{1}{2} k^{p:q} k_{[p;q]}\right]^{1/2}$, $\theta = \xi + \bar \xi =  \frac{1}{2} k^{\color{brown}l}{}_{;l}$, and $|\sigma| = \frac{1}{\sqrt{8}}\gamma = \left[ \frac{1}{2} k^{p:q} k_{(p;q)} -\theta^2\right]^{1/2}$. Sachs has provided insight into the background and motivation for research on the behavior of null congruences,\footnote{Interview by Tevian Dray with Sachs \cite{Dray:2019}.} ``Ehlers taught me general relativistic hydrodynamics. So you take a few
derivatives of the average velocity. And then you decompose it in its space
parts and timelike parts, you do an irreducible decomposition, and lo and
behold, each little part has an elegant and clear interpretation. So I think it
was my idea, why not do that for lightlike vector fields? And it doesn’t really
work if they’re not geodesics. So you do it for lightlike geodesics, and see
what comes out. And it was amazing how much came out."

The physical significance of the optical scalars follows from the geodesic deviation equation. Local coordinates can be chosen locally such that one is $u$, an affine parameter along the null geodesic, and the remaining three label neighboring geodesics. Let $\zeta^a$ represent a connecting vector to a neighboring null geodesic. Let $\frac{D}{du}$ represent the covariant derivative along the null ray, so $\frac{D \zeta^a}{du} :=  \zeta^a_{; b} k^b = k^a{}_{;b} \xi^b$. The latter equality follows from the trivial vanishing of the Lie derivative of $\zeta^a$ with respect to $k^b$ in the chosen coordinate system. Thus $k_{a;b}$ determines the variation of the connecting vector along the congruence. Of special physical  interest is the variation in the directions orthogonal to $k^a$ and $n^a$. These components can be found readily using the completeness relation $\delta^c_a = k_a n^c + n_a k^c - m_a \bar m^c -\bar m_a  m^c$, writing $k_{a;b} = \delta^c_a k_{c;d}  \delta^d_b$. Representing the resulting components as $\bar \sigma m_a m_b + \bar \rho m_a \bar m_b + c.c.$ one finds $\rho = k_{a;b} m^a \bar m^b$ and $\bar \sigma = k_{a;b} \bar m^a \bar m^b$. Then representing the connecting vectors as $\xi^a = \xi \bar m^a + \bar \xi m^a$ one can quickly determine and interpret physically the variation of $\xi$ along the geodesic. One finds, for example, that an ellipse might undergo an alteration of its semimajor axis. This shear is measured by the ratio and is $1 + 2 |\sigma| du$.

Robinson and Trautman  commenced their long-term collaboration following Robinson's visit to Warsaw in the Spring of 1959. Robinson was for the academic year 1959-60 at the University of North Carolina while Trautman had a post-doctoral appointment at Imperial College London. During this time they interacted through written correspondence.  They were finally able to work together at Syracuse during the academic year 1960-61 - before Trautman went to King's College in September, 1961.
 From there Trautman returned to an Assistant Professorship at Warsaw.  In the meantime Robinson assumed a research position at Cornell University before returning to Syracuse in 1962. Then in 1963 he joined the newly formed Southwest Center for Advanced Studies in Dallas.

They published their first paper together in 1960 \cite{Robinson:1960aa}, a short note on spherical gravitational waves. While together at Syracuse they completed a detailed analysis of this new class of exact  solutions of Einstein's equations \cite{Robinson:1962aa}. They presented a comprehensive overview at the Jablonna-Warsaw meeting in July, 1962 \cite{Robinson:1964aa}, and we  will outline the main points of this survey. The lecture is devoted to a discussion of their published `works on gravitational fields that admit a congruence of null geodesics without shear'.\footnote{\cite{Robinson:1964aa}, p. 107} In 1962 Goldberg and Sachs proved that all algebraically special solutions of Einstein's vacuum equations are shear-free \cite{Goldberg:1962ab}. Robinson and Trautman later extended the theorem to include certain restrictions on the Ricci tensor which are weaker than the Einstein equations.

The authors begin the 1964 summary with an illustrative electromagnetic example in flat spacetime, constructed with the use of null bivectors representing a plane wave that impinges upon and is reflected from a parabolic mirror, noting that the field is regular exterior to the surface representing the mirror, but diverges when continued into the interior mirror axis. There is associated with both the incident and reflected waves a shear-free null congruence, and they refer to Sachs \cite{Sachs:1961aa} as well as his Infeld Festschrift contribution \cite{Sachs:1962ad} in describing the shear-free property of cast shadows. The special merit of the Robinson-Trautman work is that it provides a procedure for constructing the metric for shear-free geometries. Let the null geodesics be represented by $x^a(\rho)$, and choose $x^3 = \rho$ as a local coordinate with both $x^4 = \sigma$ and  the spatial coordinates $x^1 = \zeta$ and $x^2 = \eta$ constant along the rays. Then the null vector $k^a = \frac{d x^a}{d\rho} = \delta^a_3$ and $g_{33} = 0$. In addition from the geodesic condition that $k_a = g_{a3}$ is independent of $\rho$, introduce a null vector $l_a$ as a linear combination of $\rho_{,a}$ and $k_a$ and form spatial vectors $x^1_a$ and $x^2_a$ as linear combinations of $\zeta_{,a}$ and $\eta_{,a}$ which are orthogonal to each other and to  $k_a$ and $l_a$. Ultimately one finds that the metric may be written as
\beq
ds^2 = -P^2 \left[\left(d\zeta - a k_a dx^a \right)^2 + \left(d\eta - b k_a dx^a \right)^2 + 2 d \rho k_a dx^a + c k_a dx^a k_b dx^b\right],
\eeq
where $a,b,c$ and $P$ are functions of all the coordinates and the components of $k_a$ are independent of $\rho$.

There are now several possibilities. The first that they investigated were spaces with non-vanishing rotation (or twist). Kerr ultimately found in this class the rotating black hole solution which carries his name.\footnote{Kerr informed  Salisbury in December 2013 that he had first heard of the twisting
 Robinson-Trautman solution in Warsaw in 1962, and that this played an important role in his own discovery. He even speculated that had Robinson and/or Trautman tried, they could have independently discovered the rotating black hole solution. } The second was the class with non-diverging rays - the gravitational plane waves. The third type represented a new class of exact gravitational diverging waves, namely shear-free but with nonvanishing $\theta$. The corresponding line element is
\beq
ds^2 = - \rho^2 p^{-2} \left(d\zeta^2 + d\eta^2\right) + 2 d\rho d\sigma + \left( - 2 m \rho^{-1} + K - 2 H \rho \right) d\sigma^2,
\eeq
where $m = m(\sigma)$, $K = \Delta \ln p$, $H = \frac{\partial}{\partial \sigma} \ln p$, $\Delta := p^2 \left(\frac{\partial^2}{\partial \zeta^2} + \frac{\partial^2}{\partial \eta^2}\right)$, and $p = p(\zeta,\eta,\sigma)$ is subject to the condition $4 \left(\frac{\partial}{\partial \sigma} - 3 H\right) m - \Delta K = 0$.
This metric could be objectionable on physical grounds since it does possess a singular line analogous to the singularity present in the parabolic mirror field described in the introduction to the paper and elucidated above.

All of these exact solutions exhibited the same asymptotic behavior, depending on the algebraic Petrov type, that had been identified by Sachs in his linearized asymptotic solutions for systems with finite sources. This is Sachs' famous peeling theorem \cite{Sachs:1962ac}.

\section{A gravitational radiation reaction coda}

We begin this short addendum with a quotation from Subrahmanyan Chandrasekhar who is accredited as having introduced in a series of papers the so-called post- and post-post-Newtonian approximations to general relativistic hydrodynamics  \cite{Chandrasekhar:1965aa}\cite{Chandrasekhar:1969aa}\cite{Chandrasekhar:1969ab}. The fourth paper in the series \cite{Chandrasekhar:1970aa} , partially co-authored by Esposito, is devoted to radiation reaction which arises in what they call the $2 \frac{1}{2}$ post-Newtonian approximation (which is order five in the slow motion approximation). Chandrasekhar writes in the introduction, after briefly referring to the recent work by Thorne \cite{Thorne:1969ac} \cite{Thorne:1969ad} and Burke \cite{Burke:1969aa}, ``It should, however, be stated that the present paper derives its basic ideas from Trautman's \cite{Trautman:1958ac}\cite{Trautman:2002aa} discussion of this same problem in the framework of the original theory of Einstein, Infeld, and Hoffmann. In that discussion Trautman failed to get agreement with the predictions of the linearized theory of gravitation; but this disagreement arose, as we shall see, from a simple oversight. When this is corrected, Trautman's procedure (as applied and extended in this paper in the framework of hydrodynamics) becomes consistent with the predictions of the linearized theory. And to this writer it appears that Trautman's approach to this problem is the simplest and the most direct that has been devised so far."\footnote{\cite{Chandrasekhar:1970aa}, p. 154}

The `oversight' to which Chandrasekhar refers was eventually corrected by his wife Michalska-Trautman. She obtained her Ph. D. with Infeld in 1961. And it was she who eventually convinced Infeld of the reality of gravitational waves. But before discussing her work we require some more background. Asher Peres was evidently the first to systematically apply the de Donder gauge condition in conjunction with Trautman's retarded wave asymptotic conditions. His series of articles \cite{Peres:1959ad} \cite{Peres:1959ac} culminated in a highly cited treatise in which he claimed to have shown that the energy lost by radiating particles was precisely that predicted by the linearized theory \cite{Peres:1960ad}. As had been shown by Fock \cite{Fock:1959aa}, it is convenient to employ as the fundamental metric variable the densitized contravarient metric $\mf{g}^{\alpha \beta} :=  \left(-g\right)^{1/2}g^{\mu \nu} $. The de Donder condition is then $\mf{g}^{\alpha \beta}{}_{, \beta} = 0$, and the  equations take the form
\beq
\eta^{\alpha \beta} \frac{\partial}{\partial x^\alpha} \frac{\partial}{\partial x^\beta} \mf{g}^{\mu \nu} = - 16 \pi T^{\mu \nu} + \Theta^{\mu \nu}, \label{E1}
\eeq
where $\Theta^{\mu \nu}$ is quadratic in $\mf{g}^{\mu \nu} - \eta^{\mu \nu}$. Peres then defined $\mf{g}^{\mu \nu} =: \mf{h}^{\mu \nu} + \mf{s}^{\mu \nu}$ with the resulting Einstein equations
\beq
\eta^{\alpha \beta} \frac{\partial}{\partial x^\alpha} \frac{\partial}{\partial x^\beta} \mf{h}^{\mu \nu} = - 16 \pi T^{\mu \nu}, \label{ET}
\eeq
and
 \beq
\eta^{\alpha \beta} \frac{\partial}{\partial x^\alpha} \frac{\partial}{\partial x^\beta} \mf{s}^{\mu \nu} =\Theta^{\mu \nu}. \label{ETheta}
\eeq
He then carried out an EIH expansion with $ {}_0\mf{h}^{\mu \nu} = \eta^{\mu \nu}$. These equations are of course perfectly suited for constructing solutions which satisfy the retarded wave asymptotic conditions. The task that one confronted, as had been addressed by Trautman, was at each order to insure that once having begun to construct solutions at each order obeying the retarded asymptotic conditions, arrange that the higher order terms were consistent with these conditions. The appropriate retarded solutions to (\ref{ET}) were trivially expanded as
$$
\mf{h}^{\mu \nu} (\vec r,t) = - 4 \int dV' T^{\mu \nu}(\vec r', t- R/c)/R
$$
$$
= - 4 \int dV' T^{\mu \nu}(\vec r', t)/R + 4 \pi \int dV' T^{\mu \nu}_{,0}(\vec r', t) - 4 \pi (2!)^{-1}  \int dV' T^{\mu \nu}_{,00}(\vec r', t) + \ldots,
$$
where $R = \left|\vec r - \vec r'\right|$.
By comparing this expansion with the corresponding $R$-dependent terms that appeared in the near-field metric expansions it was possible to connect these terms to the radiated waves.\footnote{This process was called asymptotic matching by Burke who referred to the mathematical technique that was employed in the non-linear equations of hydrodynamics. Burke refers in \cite{Burke:1971aa} to the work of Chandrasekhar, Trautman, Peres, Infeld, and Michalska-Trautman with the remarkable critique that these authors `did not recognize that the singular nature of the problem and the manipulations needed to bring in radiation effects are based on ad hoc arguments rather than on routine techniques of singular perturbations. Further, the EIH work depends heavily on the use of `good' delta functions, despite the fact that {\it nonlinearities} are important in the problem.' \cite{Burke:1971aa}, p. 402. Trautman did share Burke's concern regarding good delta functions, and we  suspect he would catalogue the first remark as not atypical of a brash young beginner in the field.}

There is another practically unnoticed participant in this story, Joanna Ryten, who married Ivor Robinson in 1967.  She completed in 1958 a master's thesis under Plebanski's direction with the title `Equations of motion in general relativity in the post-post-Newtonian approximation.\footnote{ She has shared with us her translation into English of this work.}  Her work was a forerunner of Chandrasekhar in that she used as her material source a pressureless fluid - although ultimately she was concerned with isolated pole particles.  She carried out an EIH expansion with the intention of obtaining a Lagrangian in which the gravitational field was eliminated. She concluded with a vastly complicated Lagrangian describing isolated pole particles, although it simplified considerably for the two body problem. She did not write down the corresponding two particle equations of motion.
The thesis concludes with the remark ``Knowing the post-post-Newtonian Lagrangian for the two body problem, one should be able to compute a number of physical effects: some corrections to the known effects, as well as, possibly, discover some new effects. All this requires, however, long concrete calculations and we shall not discuss any of it here." (Ryten) Robinson completed her Ph. D. with the title `Motion and gravitational radiation' under Infeld in 1963, with results published the same year \cite{Ryten:1963aa, Ryten:1963ab}.  This was a careful followup of Peres' program, but employing the good delta function technique of Infeld and Plebanski to represent pole particles. She paid special attention to the asymptotic matching conditions described above, in particular addressing the more difficult problem of insuring that these conditions were satisfied at each order for the nonlinear terms present in the solutions of (\ref{ETheta}). In \cite{Ryten:1963aa} she confirmed for the two body system that the rate at which energy was transported by gravitational waves was precisely as predicted by the linear theory. In her acknowlegements she thanks Infeld for suggesting the problem, and writes that `special thanks are also due to Dr. Trautman for stimulating discussions.' In \cite{Ryten:1963ab} she wrote down the corresponding Fokker (particle) action principle valid to the ninth EIH order. In 1963 she obtained an appointment as a research associate at the newly formed Southwest Center for Advanced Studies in Dallas, and after a year at the University of North Carolina she returned to Dallas in 1965.

We now turn to the work of R\'o\.za Michalska-Trautman. She actually joined Trautman in Syracuse in 1961, and they were married in 1962. We will go into some detail in describing the paper that she wrote after Infeld's death but with joint authorship \cite{Infeld:1969aa}.\footnote{According to a private communication from Joanna Robinson, she shared an office in Warsaw with her, Bogdan Mielnik, and Wodzimierz Tulczyjew. }. Chandrasekhar refers to this paper in \cite{Chandrasekhar:1970aa} in a note added in proof: `Professor Trautman, to whom I sent a preprint version of this paper wrote on October 1, 1969: ``During the last years of his life, Leopold Infeld worked, together with my wife, on the problem of radiation and its connection with that of motion. Among other result they obtained the correct expression (i.e. in agreement with yours) for the lowest order radiative terms in the metric corresponding to a system of point particles. This is contained in a paper by L. Infeld and R. Trautman about to appear in the Annals of Physics. Your results are certainly more general than theirs."\footnote{ \cite{Chandrasekhar:1970aa}, p. 179} The paper begins with a footnote: ``Most of the results contained in this paper were obtained in 1967, jointly by the two authors; its main ideas are due to Leopold Infeld. The paper has been written after Infeld's death, by the second author who is solely responsible for any shortcomings in the presentation."\footnote{\cite{Infeld:1969aa}, p. 561}

Infeld and Michalska-Trautman take as their two-particle Lagrangian 
$$
L = \int d^3 x  \mf{G} + \Sigma_{A=1}^2 {}^Am \frac{d{}^As}{dt}
$$
where $\frac{d{}^As}{dt} = \left[ \left(g_{\alpha \beta}\right)_A {}^A \dot \xi^\alpha {}^A \dot \xi^\beta \right]^{1/2}$ and
$\left(g_{\alpha \beta}\right)_A$ signifies the value of the metric at the site of the $A$'th particle, with the singularity due to the particle itself removed. This meaning can be communicated using the `good' delta function, 
$$
\left(g_{\alpha \beta}\right)_A = \int d^3x \delta^3 \left(x^a - {}^A \xi^a\right)g_{\alpha \beta}\left(\vec x, t \right). 
$$
The particle positions are parameterized by the time $t$, and ${}^A \dot  \xi^\alpha$ is therefore $\left(1, {}^A \dot  \xi^a\right)$. The particle action therefore minimizes the proper distance ${}^Ads$ fixed by the metric determined by the other particle. $\mf{G}$  is the first-order Einstein Lagrangian (\ref{ELag}). Variation of the action with respect to $g_{\alpha \beta}$ yields the Einstein equations
$$
R^{\alpha \beta} - \frac{1}{2} g^{\alpha \beta} R = - 8 \pi T^{\alpha \beta},
$$
where the particle stress-energy is
$$
T^{\alpha \beta} = \frac{\partial}{\partial g_{\alpha \beta}}\Sigma_A \frac{{}^Ads}{dt} = \Sigma_A {}^Am \frac{dt}{{}^Ads} \dot \xi^\alpha \dot \xi^\beta \delta^3 \left(x^a - {}^A\xi^a\right).
$$
 In a previous paper Infeld and Michalska-Trautman had shown that the rate of gravitational energy emission $\frac{d E}{dt} := R_0$  was given by 
 $$
 R_0 = -\frac{1}{2} \Sigma_A {}^Am \frac{dt}{{}^Ads} \left(g_{\alpha \beta,0}\right)_A \dot \xi^\alpha \dot \xi^\beta
 $$
 Undertaking an EIH expansion, they found in this paper that to the tenth order, ignoring total time derivatives,
 $$
 {}_{10} R_0 =  -\frac{1}{2} \Sigma_A {}^Am \left({}_5 g_{ab,0} \dot \xi^a \dot \xi^b + 2 {}_6 g_{0a,0} \dot \xi^a + {}_7g_{00,0}\right)_A
 $$
 Employing the de Donder gauge and the resulting Einstein equations (\ref{ET}) and (\ref{ETheta})  they were careful to require the matching condition that if ${}_n s^{\mu \nu} \approx \frac{{}_na^{\mu \nu}}{r} = - \frac{1}{2 \pi r} \int d^3x\, {}_n \Lambda^{\mu \nu}$, then the $(n+1)$ field should contain the expression $-\frac{d}{dt} \left( \frac{{}_na^{\mu \nu}}{r}\right) =  - \frac{1}{2 \pi r}\frac{d}{dt}  \int d^3 x\, {}_n \Lambda^{\mu \nu}$. The resulting expression for ${}_5g_{ab}$ differed by a factor of two from that obtained by Trautman in 1958 because the contribution ${}_5s_{ab}$ had not been taken into account. There was indeed an added gravitational stress contributing to  $T_{ab}$ that was missing. And the discovery of this omission led to the eventual agreement of this fully non-linear calculation with the linearized quadrupole result!
 
 \section{Conclusions}
 
We  have endeavored in this essay to highlight the contributions of Andrzej Trautman in the deduction of equations of motion and the contributions of both he and Ivor Robinson in the development of gravitational radiation theory by placing them in their technical historical context. Trautman certainly did learn the slow motion EIH approximation directly from Infeld who had continued to seek improvements in the formalism since his initial collaboration in 1938.  Infeld himself had worked with several researchers in Canada - Schild, Pirani, and Scheidegger - who made major contributions, and all three also influenced Trautman's groundbreaking thesis work. It was, however, Infeld's polish student Plebanski who oversaw this thesis - and he expressed a discomfort that Trautman would share over Infeld's treatment of particle sources using modified Dirac delta functions. But perhaps the most unexpected connection is with Peter Bergmann and his students at Syracuse. It was Bergmann who in 1949  reminded us that the surface integral that played a fundamental role in the EIH scheme was directly traceable to the general covariance of Einstein's field equations. As we have noted, this was of course long earlier recognized by Einstein. Bergmann's had begun a plan to investigate the implications of this covariance in an eventual phase space approach to general relativity. In 1953 his former student Joshua Goldberg elucidated the precise relation of the covariance to the particle equations of motion. Goldberg employed the superpotential whose derivation by Bergmann's student Zatzkis in 1951 exploited the strong identity that arises as a direct consequence of this covariance. He himself referred to Freud, though the original work by Lichnerowicz's thesis advisor Darmois was not cited. Goldberg derived from it expressions for power radiated by gravitational waves far from compact sources. It was also Bergmann who insisted that a definition of gravitational waves should focus on the generalization from the linearized transverse transverse components of the gravitational field, a demand with which Goldberg followed up in 1955 in his demonstration that care must be taken in insuring the consistency of coordinate conditions at higher orders in the slow motion approximation. 
 
Trautman built upon all of this work in the series of publications  that appeared in the  {\it Bulletin de l'Acad\ea mie Polonaise des Science}, and which formed the basis for his 1958 series of lectures at King's College, London. One reason these reports were received with such enthusiasm and were then widely circulated was that they constituted perhaps the first comprehensive overview of the state of the art of gravitational radiation physics. But even more importantly, Trautman offered several new advances. His exploitation of the covariance identities in different contexts led to new derivations of material equations of motion. He was the first to state in covariant form the discontinuity conditions which permitted the appearance of arbitrary functions describing gravitational wave-like solutions. He showed using the Zatzkis-Goldberg superpotential that it was not possible for a compact source undergoing periodic motion radiate four-momentum. He was the first to formulate the analogue of the Sommerfeld electromagnetic outgoing radiation condition for gravitational waves. And most significantly, it can be argued that his positive mass-energy loss formula for a gravitationally radiating system is equivalent to the Bondi mass loss result, and in fact preceded Bondi.\footnote{Chru\'sciel, Jezierski, and MacCallum have argued that this should be known as the Trautman-Bondi energy \cite{Chrusciel:1998aa}. This work did of course precede Penrose's conformal treatment of infinity where one explicitly follows null geodesics to a compactified boundary. But, as these authors note, one obtains the same mass loss result. It is interesting, in fact, that Penrose recognizes in a letter dated April 25, 1962, to Engelbert Schücking and Ivor Robinson, that "Engelbert will recognize the idea as one of his own that he mentioned to me". In fact all three, including also Trautman, had worked together in Syracuse. The object of the letter was to persuade the two to join him in writing the initial paper on this subject. They did not, but they certainly shared ideas with him. Penrose did write that he was "looking forward to seeing you both in Warsaw" where he presented a talk on this topic that appeared first in a technical note \cite{Penrose:1962ac} and then in the proceedings volume \cite{Penrose:1964aa}. It is significant that another expanded version also appeared as a technical report on the same date  \cite{Penrose:1962ab}, to be later published \cite{Penrose:1963aa}. Associated also with the early 1960s Syracuse group is an article that first appeared in Bergmann's technical report \cite{Bergmann:1963ab}. It was much later published as a Golden Oldie \cite{Penrose:1980aa}.  }

As we have seen, Trautman and Robinson joined forces after the former's visit to London. Robinson brought deep insights into the properties of null geodesics and their relation to special solutions of Einstein and Einstein Maxwell theory. Especially important for the future construction of solutions was his apparent priority in formulating the shear-free condition for these congruences. It is remarkable that his intuitions did involve complex objects - before they found employment in the complex tetrad Newman-Penrose formalism and in an even more natural manner in spinorial approaches to Einstein's theory. Most pertinent for our story is the role that congruences and associated bivectors played in the Trautman Robinson collaborative construction of wave-like exact solutions in the larger class of Robinson-Trautman solutions. They used the presumed underlying symmetry to carry out algebraic constructions. 

By 1970, roughly the end of the gravitational wave era we have addressed in this essay, definitive progress had finally been made in the back-reaction problem - the matching of gravitational energy emission with particle dynamics. A major step in this direction was taken in 1959 by Peres whose use of the densitized contravariant metric in the de Donder gauge permitted an elegant use of Trautman's outgoing radiation condition. Both R\'o\.za Michalska-Trautman and Joanna (Ryten) Robinson, Infeld's doctoral students, exploited this form in a systematic near and far field matching procedure, to deduce that the careful treatment of the non-linear terms in the slow motion EIH expansion delivered the same energy output as the linear approximation (Ryten) and for two interacting point particles delivered the matching particle back reaction (R\'o\.za Michalska-Trautman). Although Burke and Thorne did at roughly the same time develop a more general systematic asymptotic matching program,it is important to note these earlier and successful matching techniques which established the fundamental result of radiation damping in binary systems. And as we have seen, Chandrasekhar, who is recognized as the master of the post-Newtonian approximation scheme, recognized Trautman's  substantial contribution.

We would like to conclude with a word about when the renaissance in general relativity, discussed by many authors, began. Introducing the concept of the relativity’s low water or ebb period, Jean Eisenstadt dated it to between 1925 and 1955, thus implying that the flood tide commenced from that later date. Focusing on the obviously related concept of Relativity’s renaissance Clifford Will sees it commencing in 1960 \cite{Will:1986aa,Will:1989aa}. He writes that after roughly 1925 ``With only a few exceptions, most work in general relativity during the next thirty-five years was devoted to abstract mathematical questions and issues of principle, and was carried out by a small group of practitioners."\footnote{\cite{Will:1986aa}, p. 11}  But as we have seen, there were certainly more than a `handful' active in just gravitational wave research prior to this year, and they were receiving increasing international funding support not only from the US Air Force through grants administered by Joshua Goldberg at the Wright Patterson Air Force base\footnote{See \cite{Goldberg:1988aa}}, but also from the German-based Frick Foundation.
Furthermore, concerning the problem of motion, Will claims that the twenty-year period prior to the middle 1960's was a time of ``relative dormancy"\footnote{\cite{Will:2011aa}, p. 5939}.  Although it is certainly true that significant astrophysical discoveries brought many more researchers into gravitational wave research in the 1960's, we  hope to have made a convincing case that prior research, undertaken in part by Trautman and Robinson in the 1950's and early 1960's with evident international antecedents, constituted a solid theoretical foundation for the thrilling period of rapid development that came immediately after. Eisenstaedt’s original metaphor permits us to note that if 1955 was the date of relativity’s slack tide, then it would be natural that the incoming tide would initially proceed very slowly up the beach. An observer would have difficulty even noticing the progress being made until the flood came racing up the strand somewhat later. In Eisenstaedt’s view the period of ebb tide extended over 30 years (from 1925 to 1955) and it would therefore be natural if the flood tide lasted at least a comparable timespan (at least based upon the typical tidal patterns of the North Atlantic coasts of Europe
and America, where the papers under discussion were written). The end of the renaissance is often dated to 1975, so if the incoming tide began in 1955 then twenty years is not
too asymmetrical a period.

Unlike the term renaissance, the golden age of black hole physics (and surely therefore the golden age of General Relativity), was celebrated as it was taking place. The term was supposedly coined by Bill Press, according to Kip Thorne’s account in his book Black Holes and Time Warps \cite{Thorne:1994ab}. Presumably with a helpful eye fixed upon future historians Press organized a conference (only open to researchers under thirty years old) in 1975 at which the golden age was officially declared to have ended. Most authorities measure the golden age as about a decade in length, beginning either with Roy Kerr’s famous paper in 1963 \cite{Kerr:1963aa} (as argued by the organizers of the August 2004 Kerr Fest conference\footnote{See http://www2.phys.canterbury.ac.nz/kerrfest/timeline.html} or in 1964 according to Thorne\footnote{\cite{Thorne:1994ab}, p. 258}. 
Arguably Will’s dating for the renaissance is influenced by the dates of the golden age itself (a period which Will lived through, having been, like Press, a student of Thorne’s), but this paper focuses on the first few years after slack tide, when the wide and unknown beach still lay uncovered before the advancing tide and a few enterprising pioneers warily initiated the first incursions of the sea upon the land.

\section{Acknowledgements}
Salisbury would like to thank both Juergen Renn  of the Max Planck Institute for the History of Science and Dennis Lehmkuhl at the University of 
Bonn for support offered him as a visiting scholar. He thanks also the European Physical Journal H for its support of his Warsaw visit with Andrzej Trautman, and of course Andrzej himself for his patient and generous responses to his queries. In addition he extends profound thanks to Joanna Robinson for granting him access to her husband's literature files. Finally, Salisbury would like to dedicate this essay to his sorely missed associates and dear friends at the University of Texas at Dallas - Istvan Ozsv\'ath, Wolfgang Rindler, and Ivor Robinson.

! LaTeX Error: File `boondox-cal.sty' not found.
submit/5858866

\bibliographystyle{plain}
\bibliography{qgrav-V19}
\end{document}